\documentclass[10pt, journal]{IEEEtran}

\usepackage[utf8]{inputenc}
\usepackage{amsmath}
\usepackage{amsfonts}

\usepackage{amsthm}

\usepackage{float}
\usepackage{hhline}

\usepackage{amssymb}
\usepackage{pifont}
\newcommand{\cmark}{\ding{51}}%
\newcommand{\xmark}{\ding{55}}%

\usepackage{graphicx}
\graphicspath{{Figures/}}
\usepackage{epstopdf}

\usepackage{epsfig}
\usepackage{url}
\usepackage{comment}
\usepackage{cite}
\usepackage{graphics}
\usepackage{soul} 
\usepackage{multirow}

\usepackage{todonotes} 

\usepackage{graphicx, array, blindtext} 

\usepackage{graphicx}

\usepackage{wrapfig}

\usepackage{algorithm}  
\usepackage{algorithmic} 
\usepackage{tabularx}

\usepackage{psfrag,wrapfig}

\usepackage{enumerate}
\usepackage{paralist} 
\usepackage{subfiles} 

\usepackage{array}
\usepackage{ragged2e}
\newcolumntype{P}[1]{>{\RaggedRight\hspace{0pt}}p{#1}}

\usepackage{multirow}
\usepackage{multicol}

\ifCLASSOPTIONcompsoc
    \usepackage[caption=false, font=normalsize, labelfont=sf, textfont=sf]{subfig}
\else
\usepackage[caption=false, font=footnotesize]{subfig}
\fi

\newcommand{\eg}{{\it e.g.,}\xspace}
\newcommand{\viz}{{\it viz.,}\xspace}

\newcommand{\etal}{{\it et~al.}\xspace}
\newcommand{\ie}{{\it i.e.,}\xspace}
\newcommand{\etc}{{\it etc.}}
\newcommand{\ci}{{\it (i) }}
\newcommand{\cii}{{\it (ii) }}
\newcommand{\ciii}{{\it (iii) }}
\newcommand{\civ}{{\it (iv) }}
\newcommand{\ca}{{\it (a) }}
\newcommand{\cb}{{\it (b) }}
\newcommand{\cc}{{\it (c) }}
\newcommand{\cd}{{\it (d) }}
\newcommand{\ce}{{\it (e) }}

\newcounter{myoptimizationproblemctr}

\usepackage{color}
\definecolor{Red}{HTML}{C0392B}
\usepackage{color}
\definecolor{Blue}{HTML}{0000FF}

\definecolor{notecolor}{rgb}{0.8,0,0} 

\begin{document}


\title{Securing Vehicle-to-Everything (V2X) Communication Platforms}

\author{
\IEEEauthorblockN{Monowar Hasan\IEEEauthorrefmark{1}, Sibin Mohan\IEEEauthorrefmark{1}, Takayuki Shimizu\IEEEauthorrefmark{2} and Hongsheng Lu\IEEEauthorrefmark{2}} \\
\IEEEauthorblockA{\IEEEauthorrefmark{1}Department of Computer Science, University of Illinois, Urbana, IL, USA} \\
\IEEEauthorblockA{\IEEEauthorrefmark{2}R\&D Info Tech Labs,
Toyota Motor North America, Mountain View, CA, USA
} \\
Email: \{\IEEEauthorrefmark{1}mhasan11,
\IEEEauthorrefmark{1}sibin\}@illinois.edu,
\{\IEEEauthorrefmark{2}takayuki.shimizu,
\IEEEauthorrefmark{2}hongsheng.lu\}@toyota.com
}



\maketitle

\thispagestyle{plain}
\pagestyle{plain}


\begin{abstract}



Modern vehicular wireless technology enables vehicles to exchange information at any time, from any place, to any network -- forms the vehicle-to-everything (V2X) communication platforms. Despite benefits, V2X applications also face great challenges to security and privacy -- a very valid concern since breaches are not uncommon in automotive communication networks and applications.
In this survey, we provide an extensive overview of V2X ecosystem. We also review main security/privacy issues, current standardization activities and existing defense mechanisms proposed within the V2X domain. We then identified semantic gaps of existing security solutions and outline possible open issues.

\end{abstract}

\section{Introduction}
\label{sec:intro}



\begin{figure}
    \centering
    \includegraphics[scale=0.5]{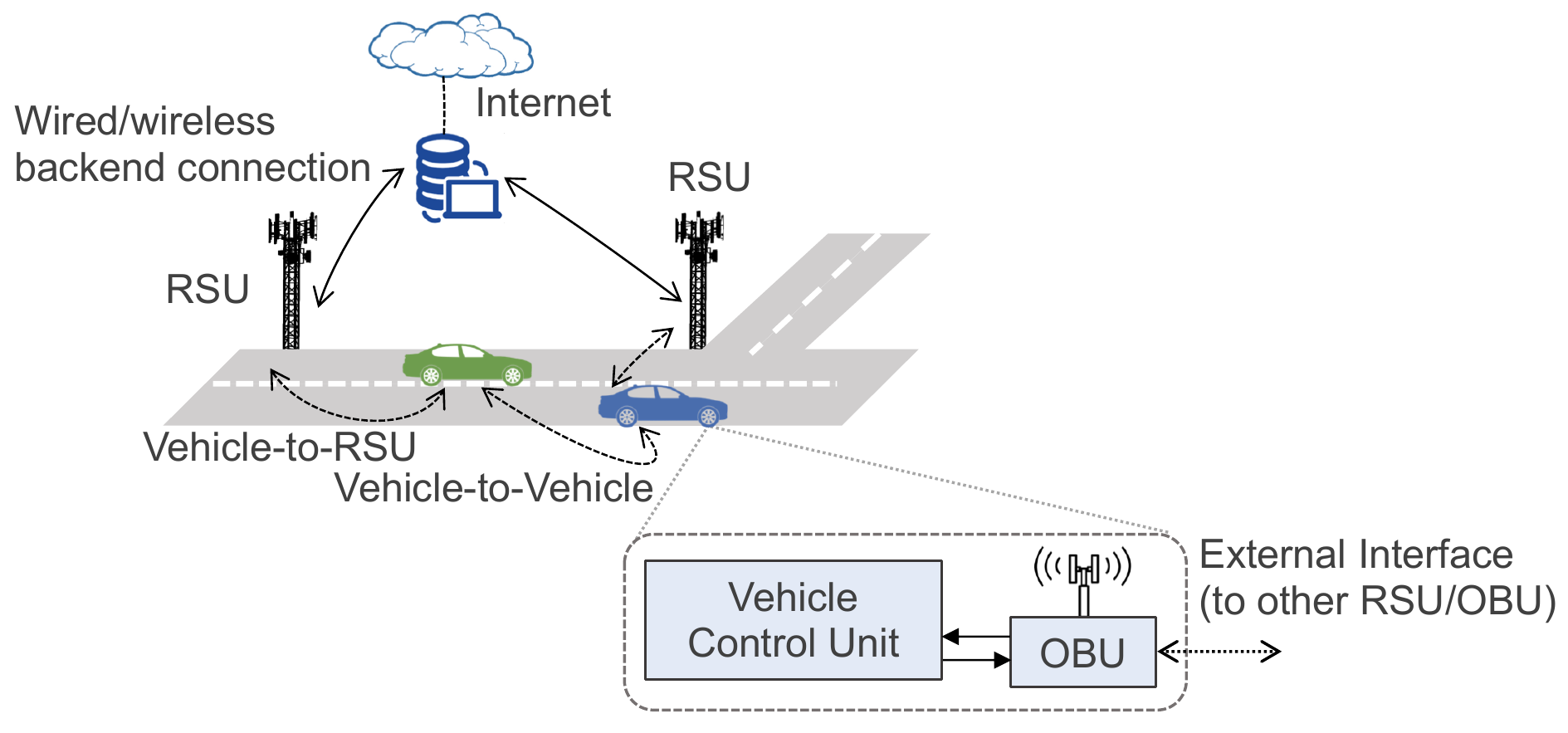}
    \caption{An illustration of V2X communication: V2X-enabled vehicles are communicating with other vehicles and infrastructures (called RSU [road-side unit]). An in-vehicle communication unit, known as on-board unit (OBU) is attached with the vehicular control system and act as an external communication interface with other entities (\eg vehicles/RSUs, \etc).}
    \label{fig:v2x_overview}
\end{figure}

Modern vehicular networks have emerged to facilitate intelligent ground transportation systems. Communication technologies in automobiles connect the various elements such as vehicles, pedestrians, infrastructures, roads, cloud computing service platforms, \etc~to each other. This has given raise to the concept of V2X (vehicle-to-everything) communications. V2X communications uses recent generation of networking technology to facilitate vehicle-to-vehicle (V2V), vehicle-to-infrastructure (V2I), vehicle-to-pedestrian (V2P) and vehicle-to-cloud (V2C) connections (see Fig.~\ref{fig:v2x_overview} for a high-level illustration). V2X communication technology is expected to improve traffic efficiency, reducing traffic incidents and road pollution, saving resources, \etc~\cite{cv2x_dsrc_comp1,s19020334}. Common use-cases for V2X applications include (but not limited to)~\cite{teixeira2014vehicular,cv2x_dsrc_comp1,5g_am2,s19020334}: road safety (\eg traffic jam/incident reporting, collision warning and collision avoidance), cooperative automated driving, infotainment services (\eg traffic information services), \etc

As with all complex connected computing platforms, extra computing capabilities in vehicles increase the exposure to potential vulnerabilities and also the likelihood of future attacks. Despite the fact that V2X communication aims to provide a
robust and resilient transportation infrastructure, V2X technologies (both existing as well as expected future developments) also pose new security challenges. For example,
a malicious vehicle can send false observation about the road (say traffic jam or an accident) and bias other vehicles to believe its incorrect observation -- as a result other vehicles are forced to change their behavior (say slow-down or reroute). Attack detection (and mitigation) is essential for widely deployed V2X systems, considering the fact that attackers may have physical access to a subset of the system. Attacks to vehicular communication systems can cause data loss, component failure and also damage environment/infrastructures.
Therefore securing V2X communicating platforms is crucial for the design, implementation and wide-scale deployment of such technology.


\subsection*{Methodology and Contributions}

While prior research\cite{parkinson2017cyber,petit2015potential} identifies some
V2X security vulnerabilities and recommends potential mitigation techniques, there is an absence of a comprehensive summary of security challenges, standardization activities and existing solutions. In this paper we investigate V2X security challenges and summarize existing solutions in a comprehensive manner.
We study over 150 papers crawled from major online literature archives (Google Scholar, IEEE Xplore, ACM Digital Library, Scopus, ScienceDirect, Wiley Online Library) published in the last 25 years (1994-2019) and identify the security issues and potential countermeasures related to V2X context. We exclude the papers that are not directly related to vehicular (communication) security domain (for instance those that are applicable to more general purpose wireless/sensor networks and/or mobile adhoc networks, \eg MANETs). We limit our study on abnormal system behavior to artifacts of malicious intent (\eg not due to hardware or component failures). We also primarily focus on the security aspects of V2X communications and provide necessary pointers for the other areas (such as radio access mechanisms, resource allocation, interference management, \etc), when needed.

In this survey we present the following contributions.

\begin{itemize}
    \item An in-depth discussion of V2X technologies and security/privacy standardization activities (Section \ref{sec:v2x_overview}).
    \item Classification and summary of potential security threats for modern V2X applications (Section~\ref{sec:v2x_threats}).
    \item A taxonomy of misbehavior detection approaches (Section \ref{sec:mis_detect}) as well as a comprehensive analysis and discussion of the state-of-the-art V2X security solutions (Sections~\ref{sec:dos_n_sybil} and \ref{sec:integrity_check}).
\end{itemize}

We also discuss possible open issues (Section~\ref{sec:discussion}), summarize multiple industry/academic/government initiatives for securing V2X communications (Section~\ref{sec:projects}) and compare our work with related surveys (Section~\ref{sec:rel_work}).

\section{V2X Platform : An Overview} \label{sec:v2x_overview}


This section provides an overview of V2X communication interfaces (Section~\ref{sec:com_interface}) and discuss various network/communication models (Section~\ref{sec:net_com_model}).

\subsection{Communication Interfaces} \label{sec:com_interface}

The internal architecture of a vehicle is interconnected with ECUs (electronic control units -- embedded computing platform that monitor/control automotive systems) coupled with sensors and actuators. The communication between the vehicle and the outside world such as other vehicles or roadside units (RSUs) is performed via external interfaces (see Fig.~\ref{fig:v2x_overview}). These vehicular external interfaces are attached to the telematics control unit (TCU) -- also referred to as \textit{on-board unit} (OBU)\footnote{In this paper we use the terms `OBU' and `vehicle' interchangeably.} -- an ECU that provides wireless connectivity~\cite{foster2015fast,den2018security}.
A vehicle control unit coordinates with the OBU to collect and disseminate vehicular data~\cite{saini2015close}.
The current standards for V2X communication are DSRC (dedicated short range communication)~\cite{dsrc_kenney2011} in the United States, C-ITS (cooperative intelligent transport systems)~\cite{cits} in Europe and ITS Connect~\cite{its_connect_itur} in Japan. Both DSRC and C-ITS operating in the 5.9 GHz ITS band while ITS Connect operating in 760 MHz band (refer to Section \ref{sec:dsrc_cits_overview} for details). An alternative to DSRC/C-ITS is the next generation of cellular wireless mobile telecommunications technology (see Section \ref{sec:cv2x_overview}). OBUs can also be equipped with interfaces for long-range communication.
These long-range wireless channels can be classified as \textit{broadcast channels} (signals can be broadcast to multiple vehicles without knowledge of the receiver's address) and \textit{addressable channels} (where messages are sent to vehicles with specific addresses.)~\cite{checkoway2011comprehensive}. Examples of broadcast channels include the global navigation satellite system (GNSS), traffic message/satellite radio receivers, \etc~Addressable channels are typically used for long-range voice/data transmissions and are intended to be used for cellular communications for mobile broadband~\cite{den2018security}.

\begin{figure*}[!h]
    \centering
    \includegraphics[scale=0.23]{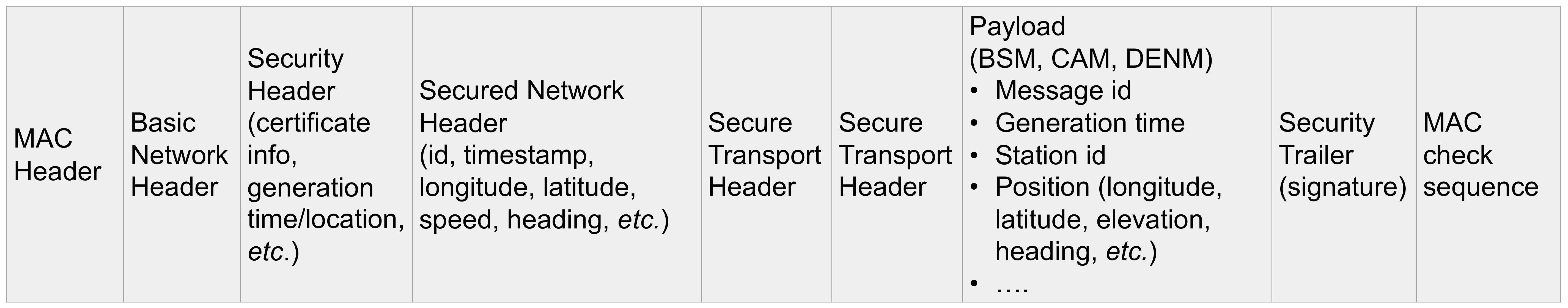}
    \caption{A high level schematic of a generic V2X packet format.}
    \label{fig:v2x_pckt}
\end{figure*}

\subsection{Network and Communication Model} \label{sec:net_com_model}

V2X communication systems~\cite{machardy2018v2x} consist of vehicle-to-vehicle
(V2V), vehicle-to-pedestrian (V2P),  vehicle-to-infrastructure (V2I), vehicle-to-cloud (V2C), vehicle-to-network (V2N) as well as vehicle-to-infrastructure-to-vehicle (V2I2V) communications. 
This can either use: \ca a technology based on IEEE 802.11p standard~\cite{IEEE802.11p} (operating in the 5.9 GHz frequency) or \cb a long-term evolution (LTE) based technology. The entities in the network can communicate with each other \ci directly (\eg using 802.11p-based technologies or LTE PC5/Sidelink interface) and/or \cii by using LTE Uu interface (uplink and downlink).


\subsubsection{IEEE 802.11p-based V2X Communications} \label{sec:dsrc_cits_overview}

As mentioned in the previous section,
the IEEE 802.11p-based adhoc V2X communication approaches are DSRC~\cite{dsrc_kenney2011} in the United States and C-ITS~\cite{cits} in Europe and ITS Connect in Japan\footnote{In Section \ref{sec:standardization}, we present the security standardization efforts in detail.}. This IEEE 802.11p-based V2X communication technology is mature and is already deployed in several countries~\cite{v2x_roadmap}.

Networking patterns for V2X communications are mainly broadcast and unicast/multicast as information messages~\cite{v2v_routing_survey} -- thus suitable for a wide range of V2X applications such as large-scale traffic optimization, cooperative cruise control, lane change warnings \etc~For certain applications (\eg over-the-air software/security credential updates,  traffic and fuel management\footnote{For instance, using signal phase and time
(SPaT) messages~\cite{spat} RSU units can inform incoming vehicles about traffic light changes (\eg
green/red) -- allowing more efficient fuel management.}, non-safety applications such as infotainment/multimedia streaming, \etc) communication
with infrastructure components, \ie via RSUs
can help in
increasing the communication range and connectivity with back-end infrastructures as well as the
Internet.

The physical
transmission (PHY) and medium access control (MAC) for both DSRC and C-ITS are same, \eg based on IEEE 802.11p amendment standards~\cite{IEEE802.11p} for vehicular networks.
ITS Connect is based on the ARIB STD-T109 standard~\cite{ARIB_STANDARD} that is similar to the IEEE 802.11p~\cite{IEEE802.11p} for PHY and MAC layers.
The technical approaches of DSRC and C-ITS have many similarities and will be the focus of this paper. As mentioned earlier, both DSRC and C-ITS
operate in the 5.9 GHz band. In the United States the communication channels range from 5.825
GHz to 5.925 GHz and the spectrum is subdivided into 10 MHz channels while the European spectrum allocation is sub-divided into several parts: \ci a 30 MHz dedicated primary frequency band for safety
and traffic efficiency applications (class A); \cii 20 MHz for
non-safety applications (class B); \ciii shared channels for radio local area networks (class C); and \civ a set of reserved channels for future use (class D). In Japan, ITS Connect operates in the 760 MHz band where 9 MHz bandwidth from 755.5 MHz to 764.5 MHz is assigned for both V2V and V2I services using ITS Connect.


In order to support V2X communication the syntax and semantics of V2X messages have been defined by standardization bodies. For DSRC, the basic safety message (BSM)~\cite{sae_J2945_1} conveys core state information
about the transmitting vehicle such as  position, dynamics, status, \etc~The BSM is a two-part message -- the first (default) part is  periodic
(sent at a rate maximum rate of 10 Hz) and the second part is event-driven (\eg for emergency braking, traffic jams, \etc) and included in the
next periodic BSM message. The C-ITS equivalent of BSM are the periodic
cooperative awareness message (CAM) and the (event-driven) decentralized environmental notification message
(DENM)~\cite{etsi_itsc}. The event-driven BSM messages are suitable for local neighborhoods (\eg single hop
broadcast) where DENMs can be used for specific
geographical areas (\eg multiple hops geocast). BSM, CAM, as well as DENM do not use encryption, \ie they are transmitted unencrypted~\cite{sae_J2945_1,etsi_itsc}.
Figure~\ref{fig:v2x_pckt} depicts a generic V2X packet format. For a detailed overview of V2X communication models, protocol stack and standardization activities we refer the reader to related work~\cite{v2x_book_2}.


\subsubsection{LTE-based V2X Communications} \label{sec:cv2x_overview}

\begin{figure}
    \centering
    \includegraphics[scale=0.48]{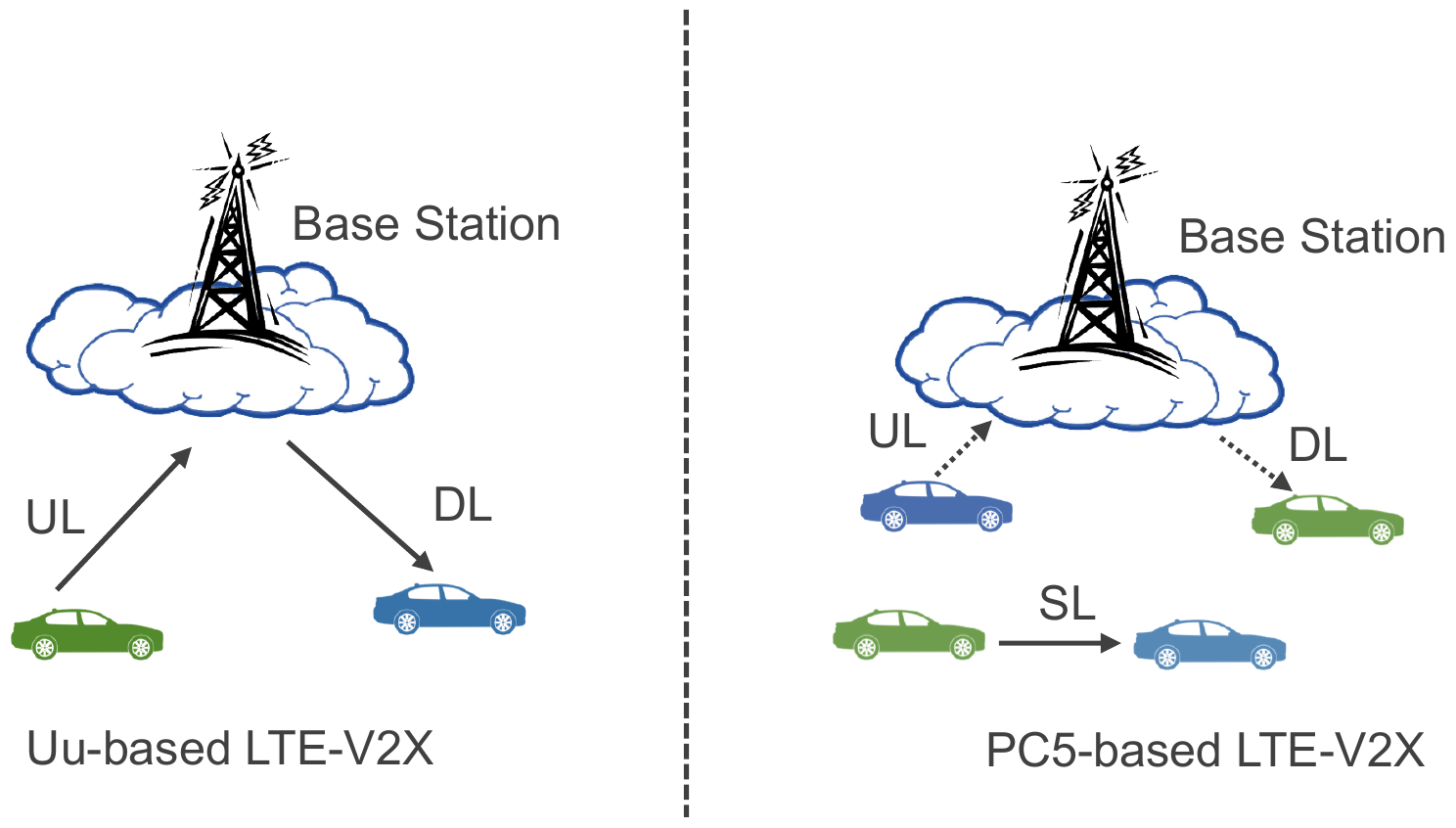}
    \caption{LTE-V2X communication modes: \ca Uu-based LTE-V2X (left): vehicles are communicating with traditional uplink (UL) and downlink (DL) channels using base station; \cb PC5-based LTE-V2X (right): vehicles use sidelinks (SL) to communicate each other with or without assistance from base stations using UL and DL for scheduling sidelink resources.
    }
    \label{fig:cv2x_overview}
\end{figure}

LTE-V2X~\cite{lte_evolution} allows vehicles to communication with each other with or without relying on base stations. 3GPP (3rd generation partnership project) Release 12 specifies proximity services (ProSe) for device-to-device (D2D) communications that enables exchange of data over short distances through a direct communication link (sidelink) based on PC5 interface (mode 1 and mode 2)
and public safety is one of the target services of LTE-D2D~\cite{lin2014overview}. LTE-V2X is an extension of 3GPP D2D functionality~\cite{hoyhtya2018review}. 3GPP Release 14 extends the ProSe functionality for LTE-V2X by using the LTE-Uu interface (uplink and downlink) and the  new PC5 interface (mode 3 and mode 4). LTE-V2X PC5 operates in the following two new modes (see Fig.~\ref{fig:cv2x_overview}): \ca  \textit{mode 3 (scheduled resource allocation mode)}: LTE-V2X PC5 mode 3 is V2X communication using sidelink with sidelink scheduling by base stations (\eg scheduling is done via Uu links);  \cb \textit{mode 4 (autonomous resource selection mode)}: LTE-V2X PC5 mode 4 is V2X communication using sidelink with autonomous sidelink resource selections by the vehicles without the help of base stations ~\cite{3gpp_rel14,5g_am2}. Both modes use PC5 for V2X communication among vehicles. In addition, mode 3 uses Uu interface for sidelink scheduling information between vehicles and base station.


 When compared to DSRC/C-ITS, the claim is that LTE-V2X systems will to provide
 larger coverage~\cite{lte_evolution, 5g_am2}. In prior work~\cite{5g_am2, 5g_am1,  wang2017_v2x_overview, cv2x_dsrc_comp1, turley2018c, filippi2017ieee, internetwork_cv2x_dsrc} researchers study and compare the adhoc V2X communications (\eg DSRC/C-ITS) with LTE-V2X in terms of radio resource allocation, performance, standardization, use-cases, deployment issues, interoperability, \etc~It is worth noting that, unlike the mature DSRC/C-ITS platforms, LTE-V2X technologies are still under development and the necessary trials/testing to
support large number of vehicles in real environments for safety applications is not yet available~\cite{filippi2017ieee, uhlemann2018battle}. In this paper, we primarily focus on the security issues for the V2X communications based on DSRC/C-ITS --  although we believe that many of those schemes can be transferred to LTE-V2X with limited (or no) modifications. We also discuss the security challenges and current solutions for LTE-V2X communication systems (Section \ref{sec:cv2x_sec}).


\section{Existing Architectures for Securing V2X Communications}


In this section we present existing cryptographic solutions for V2X security (Section~\ref{sec:pki}) and briefly discuss about various standardization efforts (Section~\ref{sec:standardization}). We mainly focus on direct V2X communication scenarios.

\subsection{Public Key Infrastructure} \label{sec:pki}

For securing V2X communications (\eg to ensure message integrity
and authenticity), the common approach is to use asymmetric cryptography using a  public key infrastructure (PKI) for the
management of security credentials~\cite{hasrouny2017vanet,vsc_pki,etsi_itsc}.
PKI enables secure exchange of messages over the network.
Each vehicle is provided an asymmetric key pair and a
certificate. The certificate contains the public key with V2X specific attributes such as ID and is signed by the key
issuing authority --
this way vehicles are registered as valid V2X participants. PKI includes the following key elements:
\ca a trusted party, \eg root certificate authority (RCA), that provides services to authenticate the identity of entities; \cb a registration authority certified by an RCA that issues certificates for specific uses permitted by
the RCA;
\cc a database that stores certificate requests and issues/revokes certificates and
\cd  an in-vehicle certificate storage  -- to save the issued certificates and private keys.

\begin{figure}
    \centering
    \includegraphics[scale=0.55]{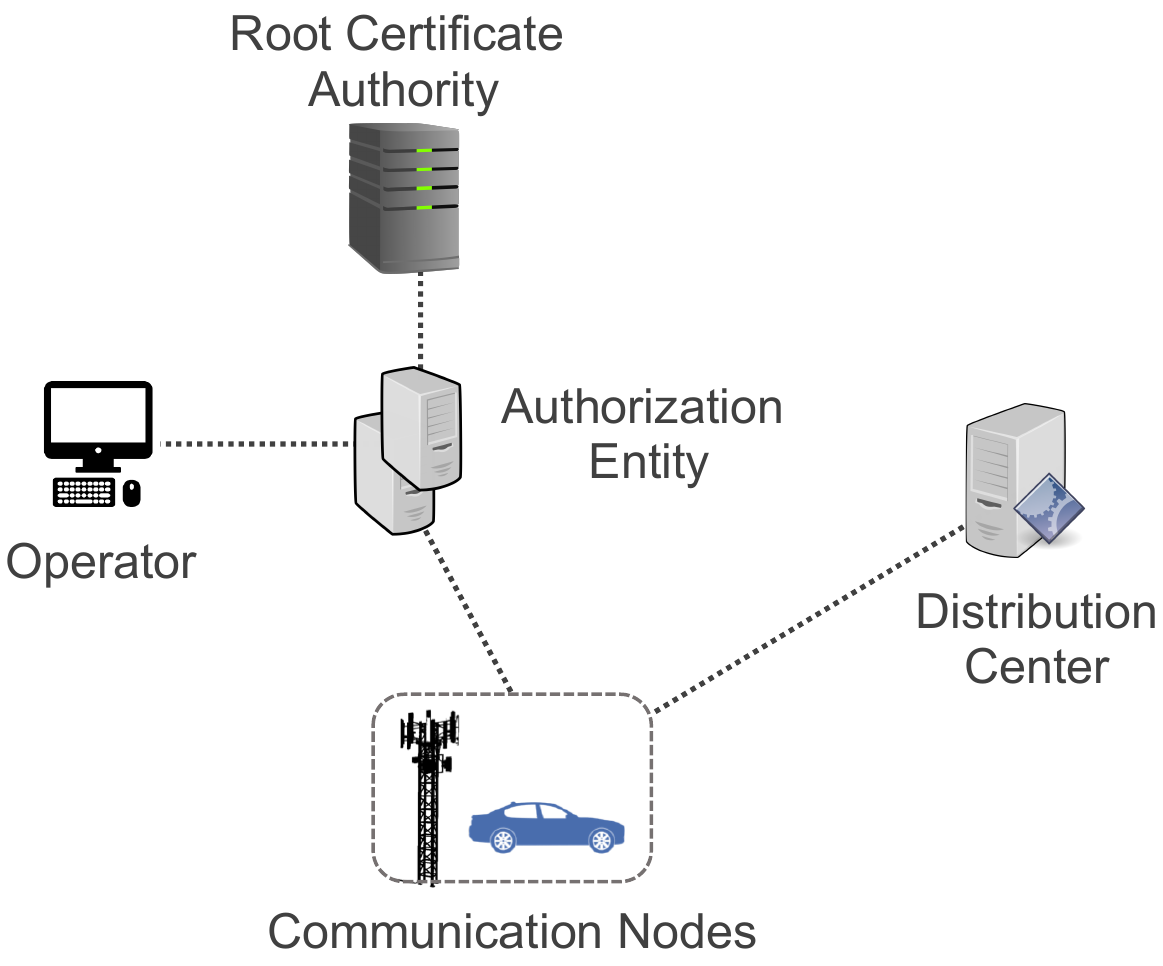}
    \caption{Schematic of a generic V2X PKI.}
    \label{fig:v2x_pki_generic}
\end{figure}

In Fig.~\ref{fig:v2x_pki_generic} we illustrate a high-level PKI for V2X communications~\cite{haidar2017performance}. The communication node (\eg vehicle and RSU) is an end-entity of the system that requests certificates
from the PKI and communicates with other end-entities. The RCA is the root
of trust for all certificates. It
delivers certificates to the authorization entities to
issue certificates to the communication nodes. The distribution center provide up-to-date trust information necessary to validate that received
information obtained from a legitimate and authorized PKI authority. The operator registers communication nodes and updates necessary information in the authorization entities.

\subsection{Standardization Efforts for V2X Security} \label{sec:standardization}


In the United States the major standardization development organizations (SDO) active in the V2X domain are IEEE
and SAE (Society of Automotive Engineers). In Europe, the relevant SDOs are
ETSI (European Telecommunications Standards Institute) and CEN (European committee for standardization). Dedicated working groups within standardization organizations and vehicle manufacturers are working on addressing security and privacy issues for V2X systems, \viz the IEEE 1609.2 working group, SAE DSRC technical committee, CAMP-VSC (crash avoidance metrics partnership--vehicle safety communications) consortium in United States and the ETSI-TC-ITS-WG5 working group in Europe addressing security and privacy issues for V2X systems\cite{harm_mdot_car,5g_am1}. Standardization groups in Europe and United States are separately building V2X security architectures based on PKI (see related work~\cite{vanet_survey} for details).

CAMP-VSC defines ``misbehavior'' as the willful or inadvertent transmission of incorrect data within the vehicular network and provides mechanisms to detect such transmissions~\cite{camp_llc_v6}. The team conceptualizes five local
misbehavior detection (LMBD) methods (to identify misbehavior within a V2V network) and three threshold-based global misbehavior detection
(GMBD) methods (identifying misbehavior at the vehicle-level using in-vehicle
algorithms and processing). The misbehavior detection techniques use a security credential management system (SCMS) and
a misbehavior authority (MA) to identify anomolous vehicles. An OBU can send misbehavior reports (MBRs) to SCMS that is based on BSM metadata, for instance: \ca the time and location where the MBR was created; \cb the LMBD method that caused the MBR creation and  \cc some combination of the start and stop time and location of the suspected misbehavior (depending on the LMBD method).



\subsubsection{V2X Security Standards}

\begin{figure}
    \centering
  \subfloat[]{%
       \includegraphics[width=3.4in]{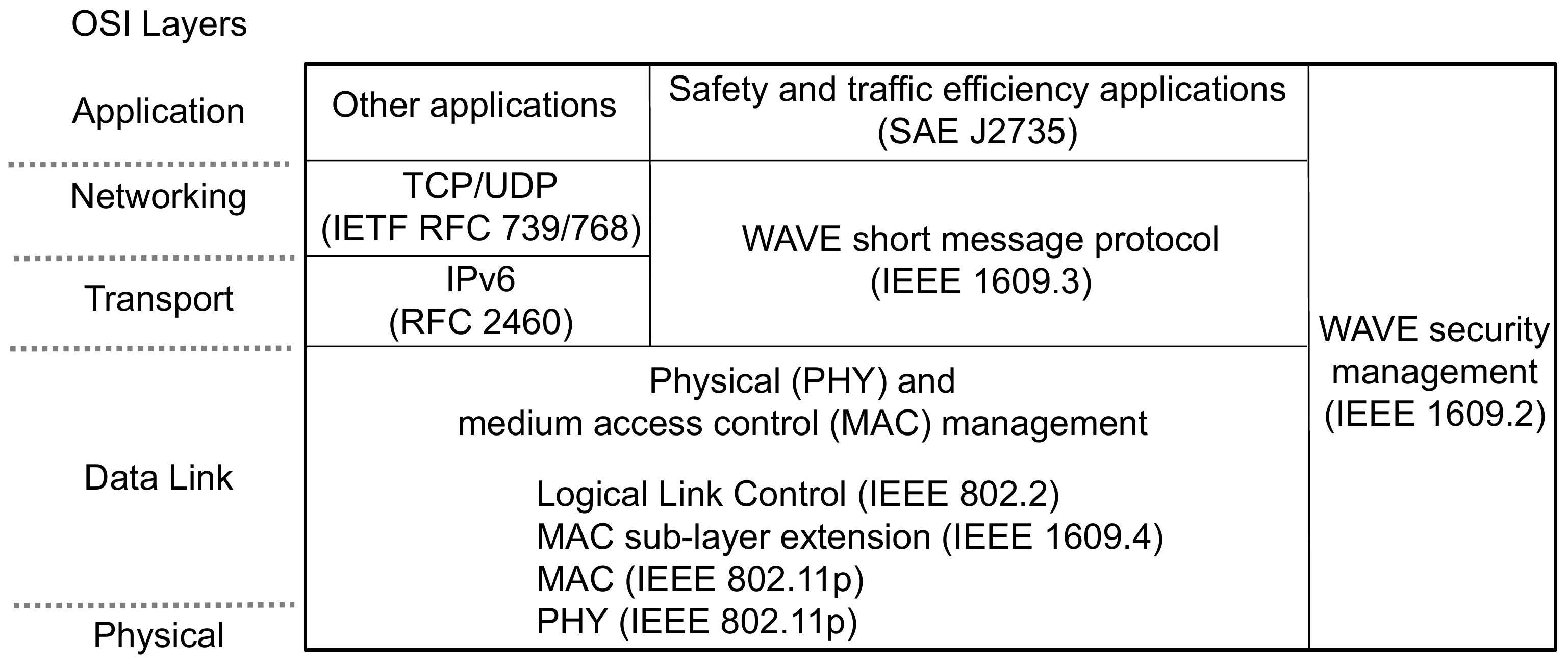}
        \label{fig:v2x_stack_us}}
    \hfill
  \subfloat[]{%
        \includegraphics[width=3.4in]{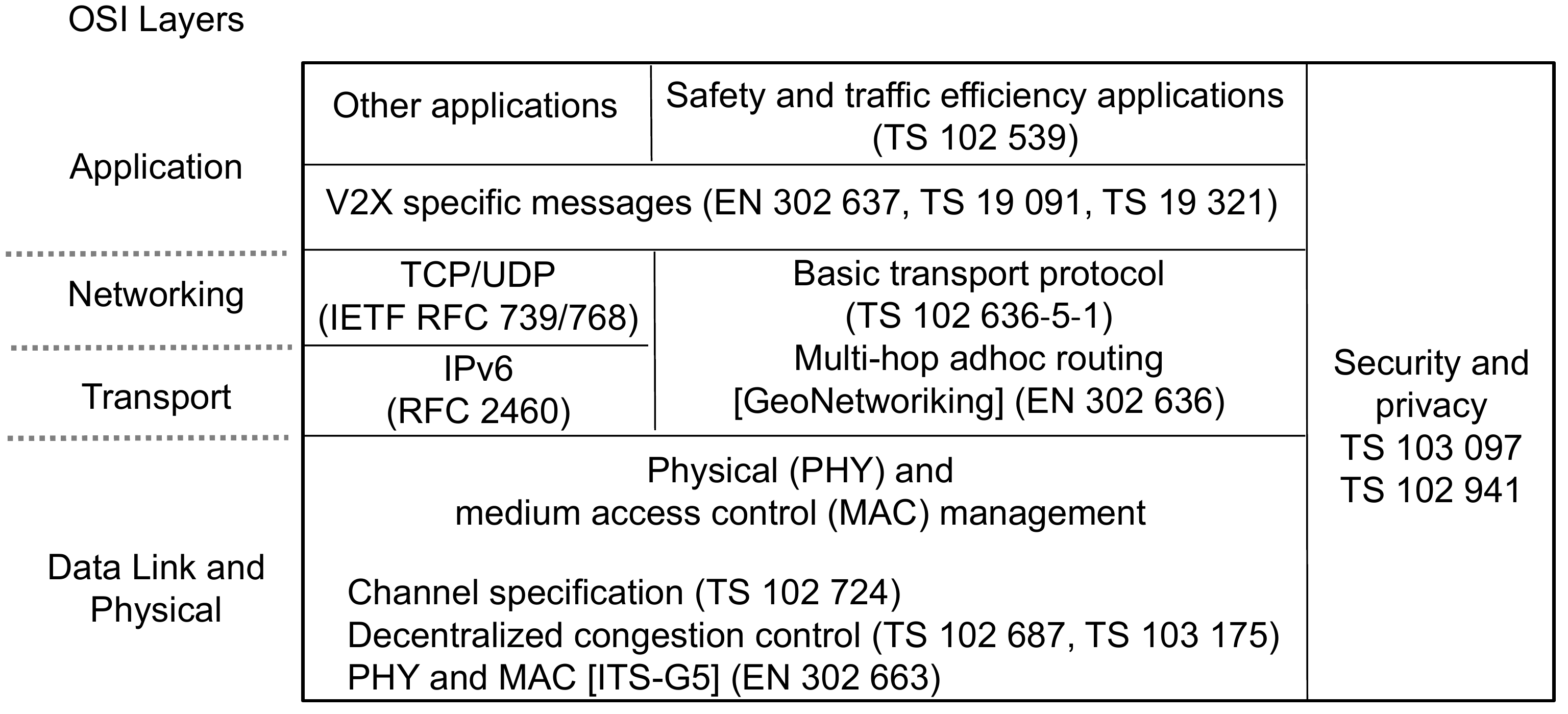}
        \label{fig:v2x_stack_europe}}
  \caption{Protocol stack and related core standards for V2X communications: \ca in United States (SAE 2945/1); \cb in Europe (ETSI-ITS).}
    \label{fig:v2x_stack}
\end{figure}

IEEE has introduced a standard for V2X communications -- WAVE (wireless access in vehicular environments)~\cite{ieee_wave, dsrc_wave_overview}. Above the protocol stack, V2X performance requirements are specified by SAE (\eg in the SAE J2945/1 standard~\cite{sae_J2945_1}) that is used primarily in the United States. ETSI has also developed standards for V2X communications, \eg ETSI-ITS  (ETSI intelligent transport system)
that includes an overall architecture, a protocol stack as well as security requirements and mechanisms~\cite{etsi_itsc}. In this section we mainly focus on the standardization of WAVE/DSRC and  ETSI-ITS since they are the most dominant technologies for actual deployment~\cite{festag2015standards}.
Figure \ref{fig:v2x_stack} depicts the protocol stacks with core networking and security standards for V2X communications in United States (Fig. \ref{fig:v2x_stack_us}) and Europe (Fig. \ref{fig:v2x_stack_europe}).

 The SAE 2945/1 standard~\cite{sae_J2945_1} uses a PKI-based
 SCMS~\cite{vsc_pki}
 for V2X security. The 
standard also requires mechanisms to protect privacy: the certificate is changed after a variable length of time and the entries in the BSM messages (that may be used to identify/track the vehicle) are randomized whenever a certificate is changed.
The V2X message
security is complaint with the IEEE 1609.2 security service standard~\cite{ieee_1609.2} that defines security
data structures, secure message formats
and the processing of those secure messages within the WAVE platform. The key features of the IEEE 1609.2 standard include: \ca wireless communication scheme between V2X devices and PKI;
\cb certificate validity and revocation schemes and \cc privacy (\eg vehicle/user identity)  preservation. For an overview of the IEEE 1609.2 standard we refer the readers to the related work~\cite[Ch. 21]{v2x_book_zhang2012}.

Security services of the IEEE 1609.2 standard support traditional cryptography mechanisms. The
service for message authenticity and integrity is based on
digital signatures. Signing and verification are performed using a public key digital signature algorithm. For instance, the sender computes a signature using the elliptic curve digital signature
algorithm (ECDSA) and the receiver verifies the signature using the associated certificate.
For transporting symmetric encryption keys, the standard uses an asymmetric encryption scheme based on the elliptic
curve integrated encryption scheme (ECIES).  The standard also defines the types of  certificate authorities (CAs), formats of the
certificates and  certificate revocation lists (CRLs).
The distribution of all security certificates (including CRLs) is performed by the SCMS.



ETSI defines architectures that applications can use to meet their security
requirements~\cite{etsi_itsc}. In order to
get access to the communication infrastructure and services, a vehicle first contacts an enrolment authority (EA) and authenticates itself. The EA replies
with a set of pseudonymous certificates (to preserve the true identity of the
vehicle as a privacy measure). These certificates validate that the vehicle can be trusted to
function correctly within the network. To request permission for accessing a service, the vehicle contacts an authorisation authority (AA) using one of the pseudonymous
certificates (that represents a temporary identity). The vehicle then receives a set of
certificates in response (one for each requested service). Vehicles can only access a service if the AA authorizes it  to use that service.

The ETSI certificate format for V2X communications is also currently based on IEEE 1609.2. The ETSI-ITS security standards were divided
into several technical reports/specifications that
describe \ca the security architecture and management, \cb trust and privacy models, \cc threat vulnerability and risk analysis, \cd messages and certificates formats and finally, \ce PKI models and mapping with IEEE 1609.2. A summary of the ETSI-ITS security standardization activities is available in earlier work~\cite{etsi_pki}. It is worth mentioning there exist services that are proposed by ETSI but not fully supported (or under development) in the SAE/IEEE (see Table \ref{tab:etsi_ieee_mapping}). A qualitative comparison of IEEE 1609.2 and ETSI-ITS standard is presented in prior research~\cite{vanet_survey, ieee_etsi_impl_comp}.

ITS Forum in Japan also provides guidelines to ensure V2X security~\cite{its_connect_sec}: \ca use of encryption (\ie chosen from the CRYPTREC~\cite{imai2000cryptrec} list of cryptographic techniques) to verify authenticity of the sender and the integrity of the messages; \cb if cryptographic keys have been leaked, the protocol must provide necessary countermeasures to minimize impact (and prevent further spread) and \cc information that stored in the vehicles/RSUs must be protected.

\begin{table}[t]
\centering
\caption{Security service compatibility in ETSI and SAE/IEEE}
\label{tab:etsi_ieee_mapping}
\begin{footnotesize}
\begin{tabular}{P{2.6cm}||P{2.5cm}|P{2.5cm}}
\hline
\bfseries Security service & \bfseries ETSI-ITS & \bfseries  IEEE 1609.2 \\ \hline \hline
Session management & By maintaining a security association & Not fully supported (on the fly association by identifying trust hierarchy) \\ \hline
Reply protection & Timestamp message and insert/validate sequence number & Timestamp message \\ \hline
Plausibility validation & Supported by data/parameter validation & Basic support (based on geographic location or message expiry time) \\ \hline
Misbehavior reporting & Not-supported & Not supported \\ 
\hline
\end{tabular}
\end{footnotesize}
\end{table}

\subsubsection{Harmonization Efforts}

There were two harmonization task groups (HTG) established by the United States and Europian international standards harmonization working group~\cite{ivanov2018cyber}: \ca HTG1 -- to harmonize security standards (\eg from CEN, ETSI and IEEE) and promote cooperative V2X interoperability; and \cb HTG3 -- to harmonize communications protocols. The goal of HTGs
was to provide feedback for SDOs and
identify areas where policy and/or regulatory actions can help to improve V2X security~\cite{harm_mdot_car}.
The harmonization efforts were completed in 2013~\cite{hg_com} and the reports/recommendations are publicly available online~\cite{hgt1_report,hgt3_report}.


\section{Security Threats in V2X Systems} \label{sec:v2x_threats}


Security threats to V2X systems depend on attacker's
capabilities and methods available to access the target (\eg vehicle, RSU and communication channels). Incentives to destabilize V2X systems include~\cite{stotz2011preserve}: \ca physical damage/vandalism (\eg denial of service, causing an accident, undesired road congestion by traffic rerouting, \etc); \cb financial incentives (\eg steal user's private information, extract OEM intellectual properties, insurance fraud, \etc); and \cc non-monetary (\eg enhancement of attackers traffic conditions,  improved attacker reputation, \etc).

\subsection{Attack Variants}

We first enumerate the attacker models that are used in the literature~\cite{raya2007securing, bissmeyer2014misbehavior,its_tvra}.
Various attacks to V2X systems can be \textit{active} or \textit{passive} -- in the case of active attacks, the adversary actively interacts with the system while the passive attackers would eavesdrop on critical data (such as private key, certificates, sensor information, \etc) without directly interacting with the system and/or disrupting normal behavior. Examples of active attacks include false code/data injection, denial-of-service (DoS), alteration of transmitted data (\eg GPS spoofing, broadcast/transaction tampering~\cite{dsrc_wave_threats}), \etc~In the V2X context, passive attacks could threaten a user's privacy since it is possible to link V2X messages and vehicle movements to individuals. Attacks can be performed \textit{offline}, \eg when the system is not operational -- these types of attacks often require physical access to the device. \textit{Online} attacks, in contrast, can be performed by exploiting hardware/software/communication bugs at runtime. The attacker could be: \ci an authenticated member of the network
allowed to communicate with other members and/or has system-level access\footnote{Not necessarily physical access to the system.} (\textit{internal}) -- these attackers behave according to the underlying protocol but send false/tampered information or \cii may not have valid credential/system access (\textit{external}) -- rather passively eavesdrop on the communication to infer sensitive information.


\subsection{V2X Attack Classifications} \label{sec:attacks_def}

\begin{figure*}
    \centering
  \subfloat[]{%
       \includegraphics[scale=0.5]{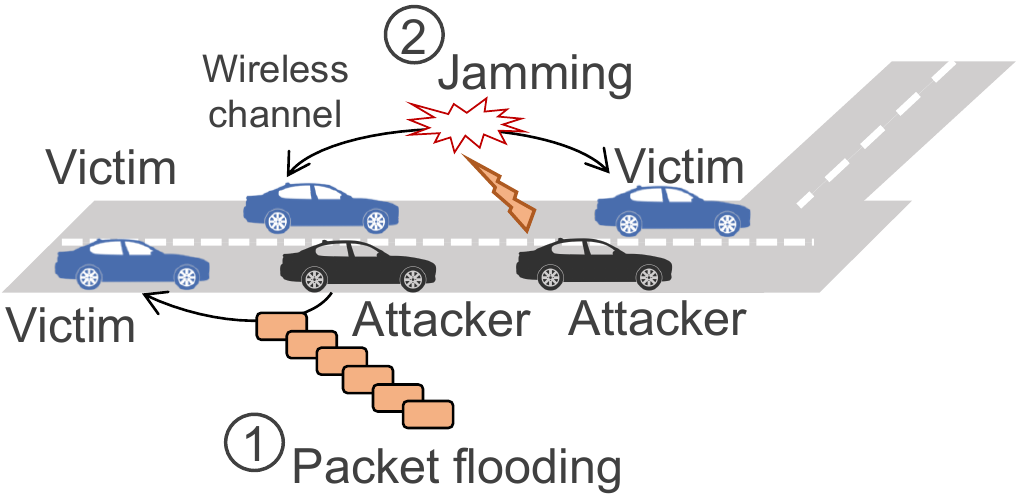}
        \label{fig:attack_dos}}
    \subfloat[]{%
        \includegraphics[scale=0.5]{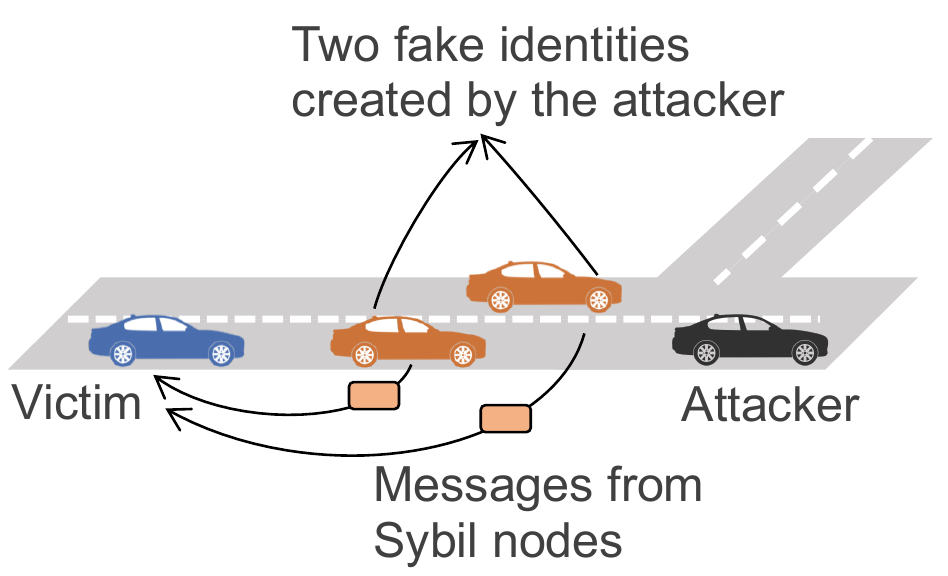}
        \label{fig:attack_sybil}}
        \hspace*{2em}
  \subfloat[]{%
        \includegraphics[scale=0.5]{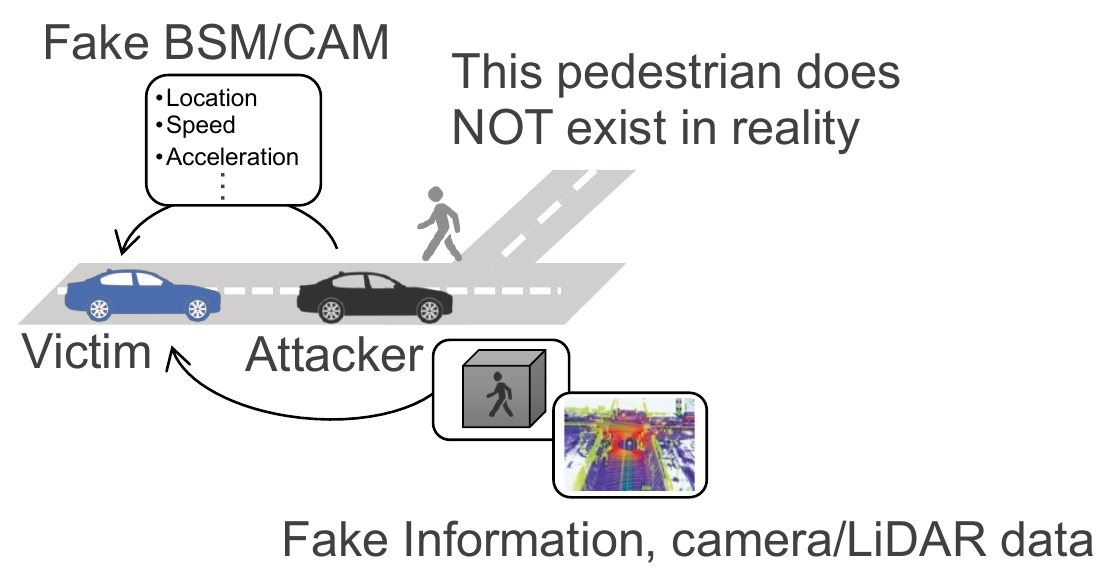}
       \label{fig:attack_falseinfo}}
  \caption{Possible attacks scenarios of V2X with malicious (black) and victim (blue) vehicles: \ca DoS attacks: \textcircled{1}  attacker floods message packets and  \textcircled{2} jams the communication channel; \cb Sybil attacks: adversary creates two fake identities and send false messages; \cc false data injection: attacker sends incorrect information (\eg about location, sensor data, object/pedestrian info, \etc).}
    \label{fig:v2x_attacks}
\end{figure*}

We now briefly review potential attacks on the V2X systems (see Fig.~\ref{fig:v2x_attacks} for a high-level illustration). While there exist prior surveys~\cite{vanet_survey, vanet_iov_survey, misbhv_survey_2019} that discuss possible attacks for vehicular networks,  in this paper we primarily focus on attacks that can be performed within the scope of existing V2X security mechanisms.
In Tables \ref{tab:threat_v2x} and \ref{tab:v2x_attack_by_scn}  we summarize the possible attacks for V2X systems. The major attacks we focus on: \ca DoS (Section~\ref{sec:th_dos}), \cb Sybil (Section~\ref{sec:th_sybil}) and \cc false data injection (Section~\ref{sec:th_falsedata}).


\begin{table}[t]
\centering

\caption{Attacks in Different Communication Scenarios}

\label{tab:v2x_attack_by_scn}
\begin{footnotesize}
\newcolumntype{C}[1]{>{\centering\arraybackslash}p{#1}}
\begin{tabular}{C{1.0cm}|C{0.5cm}|C{0.5cm}|C{3.5cm}}
\hline
\multirow{2}{*}{\bfseries Attack} & \multicolumn{2}{c|}{\bfseries Communication Scenario} & \multirow{2}{*}{\bfseries Remarks} \\
\cline{2-3}
  & \multicolumn{1}{c|}{\bfseries Broadcast} & \multicolumn{1}{c|}{\bfseries Multi-hop} &   \\
\hline
\hline
Jamming & \multicolumn{1}{c|}{\cmark} & \multicolumn{1}{c|}{\xmark} & \multicolumn{1}{P{3.5cm}}{Limited by the attacker's communication range} \\
\hline
Data flooding & \multicolumn{1}{c|}{\xmark} & \multicolumn{1}{c|}{\cmark} & \multicolumn{1}{P{3.5cm}}{No routing/forwarding is involved in broadcast of BSM/CAM} \\
\hline
Sybil & \multicolumn{1}{c|}{\cmark} & \multicolumn{1}{c|}{\cmark} & \multicolumn{1}{P{3.5cm}}{Vehicles may forward wrong (DENM) messages received from Sybil node in multi-hop scenarios} \\
\hline
Message replay & \multicolumn{1}{c|}{\cmark} & \multicolumn{1}{c|}{\cmark} & \multicolumn{1}{P{3.5cm}}{Reduces network throughput especially for multi-hop scenarios} \\
\hline
\end{tabular}
\vspace{-0.8em}
\begin{flushleft}
\scriptsize
Legends: \\
\quad \ca \cmark The attack poses threats to the communication scenario and \\
\quad \cb \xmark The attack does not disrupt the communication scenario .
\end{flushleft}
\end{footnotesize}

\end{table}


\begin{table}[t]
\centering
\caption{Major Threats to V2X Systems}
\label{tab:threat_v2x}
\begin{footnotesize}
\begin{tabular}{P{2.25cm}|P{2.8cm}|P{2.55cm}}
\hline
\bfseries Attacks & \bfseries Variants & \bfseries  Network Stack \\ \hline \hline
DoS: & &  \\
\hspace{0.5em} $\bullet$ Routing-based& Active, online, internal & Network \\
\hspace{0.5em} $\bullet$ Flooding & Active, online, internal & Application, network \\
\hspace{0.5em} $\bullet$ Jamming & Active, online, external & Physical \\
\hline
Sybil & Active, online, external/internal  & Application, transport, network, data link \\
\hline
False data injection & Active, online, internal & Application, transport, network, data link, physical   \\
\hline
\end{tabular}
\end{footnotesize}
\end{table}




    \subsubsection{DoS Attacks} \label{sec:th_dos}

    DoS attacks can happen in different layers of the network where an adversary sends more requests than the system can handle. For instance, an attacker could try to shutdown/disrupt the network established by RSUs and stop communication between vehicles and/or RSUs~\cite{vanet_iov_survey}. In a distributed DoS (DDoS) attack~\cite{sumra2014effects} malicious nodes launch attacks from different locations thus making it harder to detect. In the physical layer, an important type of DoS attack is the \textit{jamming attack}~\cite{jamming2} (refer to the related work~\cite{punal2012vanets} for detailed classification) where the attacker disrupts the communication channel (\eg by electromagnatic interference) and can filter/limit incoming messages.
    Jamming functions well only in geographically restricted areas, \ie say within the range of the attacker(s) wireless device. We also note that most jamming/DoS attacks on the PHY level (IEEE 802.11p) or  the bands around 5.9 GHz are always restricted by the range of the attacker(s) and do not impact V2X communications everywhere.
    Jamming attack does not require any particular knowledge of the semantics of the exchanged messages~\cite{misbhv_survey_2019}. Although jamming attacks are not specific to V2X systems (\ie can be a threat for any wireless network), such attacks can increase the latency in the V2X communications and reduce the reliability of the network~\cite{punal2015experimental}. 

    In the network layer, routing-based DoS attacks such as
    \textit{JellyFish attack}~\cite{jellyfish} exploits vulnerabilities in congestion control protocols and the attacker delays or (periodically) drops packets (albeit does not violate protocol specifications). Packet dropping is catastrophic for safety-related applications -- for instance, a vehicle involved in a traffic accident should propagate warning messages, but other vehicles could be prevented from receiving these warning messages by an attacker who intentionally drops/miss-routes packets. Another variant is the \textit{intelligent cheater attack}~\cite{vanet_iov_survey} where an adversary obeys the routing protocol specifications but misbehaves intermittently. Such attacks require long term monitoring for detection~\cite{jellyfish} that could be impracticable for V2X scenario due to high mobility. \textit{Flooding attacks}~\cite{vanet_iov_survey} such as data flooding (\eg where an attacker creates bogus data packets and sends it to their neighbors)
    can make the network resources (\eg bandwidth, power, \etc) unavailable to legitimate users. We note that these routing-based attacks can only be performed to multi-hop communication networks (\eg not single-hop direct communications such as broadcasting BSM).

    \subsubsection{Sybil Attacks} \label{sec:th_sybil}


    This is a well-known harmful attack in wireless vehicular networks where a vehicle pretends to have more than one identify (\eg multiple certified key-pairs) either at the same time or in succession~\cite{yu2013detecting}. Sybil attackers may also launch DoS attacks, waste network bandwidth, destabilize the overall network and pose threats to  safety~\cite{vanet_iov_survey,local_coop_survey18}. For instance, if a malicious vehicle changes its identity,
    it may use multiple pseudonyms to appear as a different, moving vehicle or make it appear that the road is congested (even though it is not)
    and send incorrect information about the road conditions to neighbouring vehicles/RSUs. A Sybil attacker could also use the pseudo-identities to maliciously boost the reputation/trust score (\eg that use to measure how much neighbors can rely on information send by a given vehicle $V_i$), \etc~of specific vehicles or, conversely, reduce the score of legitimate vehicles~\cite{misbhv_survey_2019}.

    \subsubsection{False Data Injection} \label{sec:th_falsedata}


    A rogue vehicle could generate false traffic/safety messages or incorrect traffic estimation information (that differs from real-world information) and broadcast it to the network with the intention of disrupting road traffic or triggering a collision~\cite{verific_survey, vanet_iov_survey}.
    Sybil attackers can claim their existence at multiple locations and can thus inject false information in the network. By GPS spoofing
    an attacker could inject false position information by using GPS simulators and the victim vehicles may end up accepting these generated, fake, (but stronger than original) signals.  Incorrect data such as falsified location information could decrease message delivery efficiency by up to approximately 90\%~\cite{leinmuller2006greedy}. Researchers have shown that cooperative adaptive cruise control (CACC) -- an important V2X use-case -- is specifically vulnerable to false data injection attacks~\cite{cacc2, cacc1}.



    Another type of false data injection is \textit{replay attack} where an attacker re-transmits messages to exploit the conditions at the time when the original message was sent (\eg the attacker stores the event information and will resend it later, even though it is no longer valid)~\cite{hasrouny2017vanet,vanet_iov_survey}. For instance, in location-based replay attacks
    the attacker records an authenticated message at a location $L_i$, transmits it quickly to a location $L_j$ (and re-broadcasts it at $L_j$). Similarly, in time-based replay attacks, an adversary records a valid message at time $t_1$ and replays it later (at the same location) at another time $t_2$. For replay protection, there exist mechanisms such as: \ca including a time stamp in every message -- say by using a global navigation satellite system (GNSS)~\cite{gnss_survey_2016} and/or \cb digitally signing and including sequence numbers~\cite{dsrc_wave_overview, etsi_pki}, \etc~The V2X standards~\cite{ieee_wave,etsi_itsc} also provide mechanisms for replay protection: the maximum transmission delay of single-hop messages would need to be verified by receiving stations and messages with an outdated timestamp (or a future timestamp) should be considered as not plausible. Replay attacks in multi-hop V2X communication (\ie DENMs) are related to routing misbehavior (\eg where the attacker may deviate from the routing protocol and reroute messages to specific vehicles and/or drop messages)~\cite{vanet_survey, vanet_iov_survey}.
    While replay attacks (specially for multi-hop communications) can affect network throughput, support of infrastructures such as RSUs (and base stations for C-V2X) can reduce the impact of routing misbehavior~\cite{misbhv_survey_2019}.



\section{Misbehavior in V2X Communications} \label{sec:mis_detect}

In V2X security literature researchers often use the term \textit{misbehavior}~\cite{raya2007securing,raya2007eviction,verific_survey,vanet_iov_survey,misbhv_survey_2019}. This commonly refers to attacks that are executed by the malicious entity, \eg a \textit{misbehaving node} transmits erroneous data that it should not transmit when the system is behaving as expected. This is different than \textit{faulty nodes}~\cite{leinmuller2007security,lin2008security,kargl2008secure}, \ie when an entity produces incorrect or inaccurate data without malicious intent. While  these definitions are not consistently used in literature, in this paper we use `misbehavior detection' to refer to the uncovering of malicious entities.

In the following, we \ca describe threat model and commons assumptions on adversarial capabilities (Section~\ref{sec:threat_model} and then \cb provide a summary of various misbehavior detection mechanisms (Section~\ref{sec:mis_dis_tax}).

\subsection{Threat Model} \label{sec:threat_model}

As we discussed in Section \ref{sec:v2x_overview}, protection of wireless V2X communications by use of cryptographic credentials is a common approach. In the following we assume that the attacker has the credentials to communicate with other vehicles in the network (\eg an internal attacker) 
and the attackers
are able to distribute bogus information~\cite{ruj2011data}. For instance, the attacker could  send false information or conceal some information, tamper with its own message contents (\eg event type/location, node position, \etc), generate false messages or bias another vehicle's decisions (by sending erroneous messages).
We also assume that the RSUs are trusted in general (although in Section~\ref{sec:discussion} we discuss cases when we relax this assumption).

\subsection{Classification of Detection and Prevention Mechanisms} \label{sec:mis_dis_tax}

\begin{figure}
    \centering
    \includegraphics[width=3.6in]{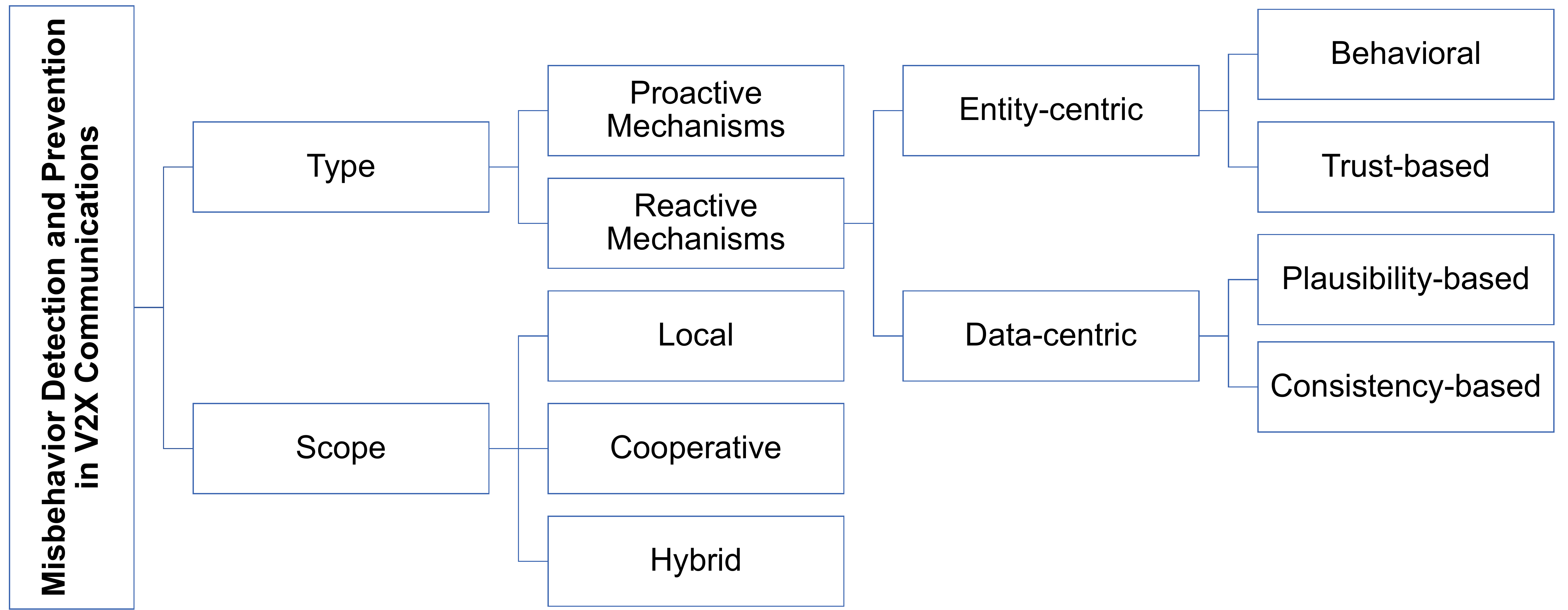}
    \caption{Taxonomy of V2X misbehavior detection/prevention approaches. In Table~\ref{tab:attack_v2x} we summarize how various classes of attacks can be detected by these approaches.}
    \label{fig:v2x_a_cls}
\end{figure}

In Fig.~\ref{fig:v2x_a_cls} we illustrate various misbehavior detection/prevention approaches. V2X security approaches can broadly be characterized as \textit{proactive} and \textit{reactive} mechanisms~\cite{verific_survey, misbhv_survey_2019}. Proactive security refers to any kind of mechanism that enforces a security policy -- say use of a PKI, digital signatures and certificates, tamper-proof hardware, \etc~ This reduces the chances of bogus information exchange by unauthorized entities due to lack of credentials and can be maintained through a combination of infrastructure and tamper-proof hardware~\cite{karagiannis2011vehicular}. While these mechanisms reduce attack surfaces by detracting external attackers, insider attackers can generate legitimate false information. Such schemes also face scalability and complex management issues (\eg key management, revocation, trust establishment in multi-hop communication). Reactive mechanisms can be enforced where the attacks cannot be prevented by proactive security policies. These mechanisms can be grouped into two classes: 
\ca \textit{entity-centric} and \cb \textit{data-centric}. Entity-centric approaches focus on identifying the misbehaving node generally based on trust establishment (say from past behavior/interactions) by using a PKI or in a cooperative manner (\eg using signature verification). Data-centric approaches, in contrast, verify the correctness of the received data (instead of investigating the trustworthiness of the sender).

Entity-centric detection approaches can be further subdivided into: \ca \textit{behavioral} (\eg observes patterns in the behavior of specific nodes at the protocol level) and \cb \textit{trust-based} (\eg evaluation of trust-score, often using a central authority to remove malicious nodes). Data-centric mechanisms are similar to intrusion detection in traditional computing systems that correlate the received information with the information already known from previous history/behavior. These approaches can be either: \ca \textit{plausibility-based} (model-based approach that verifies if the information transmitted from a particular sender is consistent with the model) or \cb \textit{consistency-based} (\eg use information of packets -- generally from multiple participants -- to determine the trustworthiness of new data). We highlight that entity-centric and data-centric detection mechanisms are mostly orthogonal and often researchers propose to use combinations of both types.
Depending on the scope, detection mechanisms can be: \ca \textit{local} (\eg performed locally, say by vehicle OBUs and not affected by detection mechanisms in other vehicles; or in the back-end by the RSUs) and/or \cb \textit{cooperative} (detection relies on collaboration between vehicles/RSUs). In contrast to RSU-based mechanisms, OBU-based approaches do not need dedicated infrastructure (\eg vehicles performing situation evaluation by themselves without any infrastructure). Researchers also proposed hybrid approaches where both RSU and OBUs are jointly involved in misbehavior detection (see Sections~\ref{sec:dos_n_sybil} and \ref{sec:integrity_check}). Behavioral and plausibility schemes generally operate locally while consistency and trust-based rely on cooperation among vehicles/RSUs to detect inconsistencies. Some consistency-based mechanisms can also be performed locally for more fine-grained detection with the cost of exposing them to Sybil attacks.


We now briefly review the mechanisms to secure V2X communications from different classes of attacks (Sections~\ref{sec:dos_n_sybil} and \ref{sec:integrity_check}). Table \ref{tab:attack_v2x} summarizes the exiting solutions.

\begin{table*}[t]
\centering
\caption{Summary of Misbehavior Detection Mechanisms for V2X Communications\textsuperscript{*}}
\label{tab:attack_v2x}
\begin{footnotesize}
\begin{flushleft}
\scriptsize
\textsuperscript{*}The terms OBU-L, OBU-C and OBU-L/C represent whether a vehicle OBU performs decisions: \ca locally (OBU-L), \cb with the involvement of neighboring vehicles (OBU-C) or \cc using the combination of both (OBU-L/C). If any scheme requires assistance of infrastructure (CA, MA, \etc~say for misbehavior reporting), we consider RSUs as the entry point to communicate with the back-end.
\end{flushleft}

\begin{tabular}{P{2.3cm}|P{1.7cm}|P{1.6cm}|P{2.00cm}|P{4.5cm}|P{3.5cm}}
\hline
\bfseries Reference & \bfseries Decision Involvement & \bfseries Approach & \bfseries Defense Against & \bfseries Key Idea & \bfseries Major Limitation(s) \\ \hline \hline
Hamieh \etal~\cite{hamieh2009detection},  Lyamin \etal~\cite{lyamin2018ai} & OBU-L  & Entity-centric & DoS (jamming) & Detect patterns in radio interference to differentiate jamming  and legitimate scenarios & No attacker identification\\
\hline
Chuang \etal~\cite{chuang2014team}& OBU-C, RSU & Entity-centric & DoS (flooding, resource exhaustion) & Use short-term key-pairs & Extra message exchange overhead \\
\hline
He \etal~\cite{he2012mitigating}, Studer \etal~\cite{studer2009flexible} &  RSU & Entity-centric & DoS (resource exhaustion) & Use pre-authentication~\cite{he2012mitigating}, alternative authentication mechanism with reduced memory/computation requirement~\cite{studer2009flexible} & Increased communication latency \\
\hline
Hortelano \etal~\cite{watchdog_icc10}, Daeinabi \etal~\cite{daeinabi2013detection} & OBU-L ~\cite{watchdog_icc10}, OBU-C and RSU~\cite{daeinabi2013detection} & Entity-centric & DoS (malicious packet dropping/forwarding) & Predict the (expected) behavior of the neighbors by using a watchdog & No privacy discussion and only detects malicious packet forwarding \\
\hline
Soryal \etal~\cite{6799761}, Verma \etal~\cite{verma2013prevention,verma2015bloom} & RSU & Entity-centric & DoS (packet flooding)  & Monitor message exchange pattern & Lack of scalability \\
\hline
Biswas \etal~\cite{biswas2012ddos} & RSU &  Entity-centric & DDoS (packet flooding) & Randomize message schedule of the RSU & Does not work if the attacker can reproduce the randomized schedule\\
\hline
Hasrouny \etal~\cite{hasrouny2017security}  & OBU-L/C, RSU & Entity-centric & DoS (packet flooding) & Limit the number of accepted received messages by calculating a trust score & Lack of implementation details and performance evaluation \\
\hline
Kerrache \etal~\cite{kerrache2017tfdd} & OBU-C, RSU & Entity and data-centric & DoS (packet flooding) & Trust-based data verification and routing mechanism to eliminate misbehaving vehicles & Stealthy attacker can bypass the detection mechanisms \\
\hline \multicolumn{6}{r}{\textit{Continued on next page.}} \\
\end{tabular}
\end{footnotesize}
\end{table*}

\begin{table*}[t]
\addtocounter{table}{0}
\centering
\tablename\ \thetable\ -- \textit{Continued from previous page} \\
\vspace*{1.0em}
\begin{footnotesize}
\begin{tabular}{P{2.3cm}|P{1.7cm}|P{1.6cm}|P{2.00cm}|P{4.5cm}|P{3.5cm}}
\hline
\bfseries Reference & \bfseries Decision Involvement & \bfseries Approach & \bfseries Defense Against & \bfseries Key Idea & \bfseries Major Limitation(s) \\ \hline \hline
Grover \etal~\cite{grover2011sybil} & OBU-C & Data-centric & Sybil attack & Compare neighbor-set of several vehicles over time & Possible false-positive errors, no prevention for short-term Sybil attacks \\
\hline
Hao \etal~\cite{hao2011cooperative} & OBU-C & Data-centric & Sybil attack & Use correlation of mobility traces to identify Sybil nodes & May not detect multiple Sybil attackers \\
\hline
Golle \etal~\cite{Golle:2004} & OBU-C  & Data-centric & Sybil attack & Compare revived data with an expected model & Require a suitable model to compare against, no validations or performance test is performed \\
\hline
Guette \etal~\cite{guette2007sybil}, Yao \etal~\cite{yao2019multi} & OBU-L & Data-centric & Sybil attack & Link Sybil nodes through physical characteristics (\eg RSSI) & Infeasible with GPS errors~\cite{guette2007sybil}, could be failed if attacker uses multiple radios~\cite{yao2019multi}  \\
\hline
Ruj \etal~\cite{ruj2011data} & OBU-L, RSU & Data-centric & Sybil attack, position cheating & Monitor and compare the messages with an expected behavioral model to analyze if such events are actually happened & No validations or performance test, unrealistic assumptions \\
\hline
Xiao \etal~\cite{xiao2006detection} & OBU-C, RSU & Entity-centric & Sybil attack & Analyze signal strengths of the received beacons & No way to verify behavior of the `verifier' vehicle \\
\hline
Sowattana \etal~\cite{sowattana2017distributed} & OBU-C & Entity-centric & Sybil attack & Analyze validity of received beacons and generates
trust score (of a given vehicle) through a voting mechanism & Voting mechanism itself could be vulnerable to Sybil attack \\
\hline
Park \etal~\cite{park2009defense} & OBU-L, RSU & Entity-centric  & Sybil attack & Observe similarity of motion
trajectories & May not detect correctly for all scenarios (\eg when two vehicles coming from opposite directions)\\
\hline
Zhou \etal \cite{zhou2011p2dap} & RSU & Entity-centric & Sybil attack & Link pseudonyms to common
values using hash functions & No privacy preservation for the central authority \\
\hline
Hamed \etal~\cite{hamed2018sybil} & RSU & Entity-centric & Sybil attack & Monitor mobility pattern of the vehicles & Finding proper detection threshold, increase false positive/negative errors \\
\hline
Chen \etal~\cite{chen2009robust} & OBU-C, RSU & Entity-centric & Sybil attack & Exchange digital signature (that is periodically issued by the RSU) among neighbors and compare it with a reference trajectory & Prone to false positive errors, high overhead, does not consider vehicle privacy\\
\hline
Chang \etal~\cite{chang2012footprint} & OBU-C, RSU & Entity-centric & Sybil attack & Observe similarity of motion
trajectories & Potential false positives, geared towards urban networks only  \\
\hline
Lee \etal \cite{lee2013dtsa}, Rahbari \etal~\cite{rahbari2011efficient} & RSU & Entity-centric & Sybil attack & Use session key-based certificates\cite{lee2013dtsa} or PKI/CA to compare reply messages received from RSU~\cite{rahbari2011efficient}  & Extra information exchange overhead (\ie reduced throughput) \\
\hline
Feng \etal~\cite{feng2017method}  & OBU-C, RSU & Entity-centric & Sybil attack & Use short-term  public key and pseudonyms & Could collapse if RSUs/OBUs are compromised \\
\hline
Chen \etal~\cite{chen2011threshold}, Singh \etal~\cite{Singh:2017:RUA} & OBU-C~\cite{chen2011threshold}, OBU-L and RSU~\cite{Singh:2017:RUA} & Entity-centric & Sybil attack &  Use of specific certificates and cryptographically protected usage restriction of the credentials & Extra computation/communication overhead \\
\hline \multicolumn{6}{r}{\textit{Continued on next page.}} \\
\end{tabular}
\end{footnotesize}
\end{table*}

\begin{table*}[t]
\addtocounter{table}{0}
\centering
\tablename\ \thetable\ -- \textit{Continued from previous page} \\
\vspace*{1.0em}
\begin{footnotesize}
\begin{tabular}{P{2.3cm}|P{1.7cm}|P{1.6cm}|P{2.00cm}|P{4.5cm}|P{3.5cm}}
\hline
\bfseries Reference & \bfseries Decision Involvement & \bfseries Approach & \bfseries Defense Against & \bfseries Key Idea & \bfseries Major Limitation(s) \\ \hline \hline
Kim \etal~\cite{kim2010vanet} & OBU-C, RSU & Entity-centric & False event notification & Filter messages based local sensor information and event specific data & Highly application specific and depends on verified positions \\
\hline
Cao \etal~\cite{cao2008proof}, Hsiao \etal~\cite{hsiao2011efficient}, Petit \etal~\cite{petit2011spoofed}  & OBU-C & Entity-centric & False event notification & Determine the correctness of
event reports through voting/consensus & Weak attacker model, prone to Sybil attacks \\
\hline
Ghosh \etal~\cite{ghosh2010detecting} & OBU-L & Data-centric & False event notification & Correlate future behavior from past events & Requires specific driver behavior for validation (potential false positive/negative errors) \\
\hline
Schmidt \etal~\cite{schmidt2008vehicle} & OBU-L/C & Data-centric & False data injection & Build reputation through evaluating vehicle behavior & Assumes a honest majority, no performance test \\
\hline
Yang \etal~\cite{yang2013misdis} & OBU-L, RSU & Entity-centric & False data injection & Monitor vehicle behaviors by logging message transmissions & Requires strong authentication and identification mechanism \\
\hline
Lo \etal~\cite{lo2007illusion} & OBU-L & Data-centric & False data injection & Monitor sensor values/received messages and ensure validity using a rule database & Shared rule database can leak information to the attackers \\
\hline
Vora \etal~\cite{vora2006secure} & RSU & Entity-centric & Position cheating & Perform position verification using multiple `verifier' RSUs & Vulnerable to targeted attacks  \\
\hline
Yan \etal~\cite{yan2008providing} & OBU-C & Data-centric & Position cheating & Check consistency of messages from multiple sources (\eg on-board radar, incoming traffic data, \etc)  & Performance overhead, lack of privacy \\
\hline
St{\"u}bing \etal~\cite{stubing2010verifying,  stubing2011two}, Jaeger \etal~\cite{jaeger2012novel} & OBU-L & Data-centric & Position cheating & Analyze CAM message sequences and track the vehicles & Computationally expensive, chances of (partial) privacy breach \\
\hline
Sun \etal~\cite{8228654}, Hubaux \etal~\cite{hubaux2004security}& OBU-L~\cite{8228654}, RSU~\cite{hubaux2004security} & Data-centric & Position cheating & Verify position using physical properties (\eg Doppler speed measurements~~\cite{8228654}, speed of light~\cite{hubaux2004security}, \etc) & Additional hardware requirement~\cite{8228654}, potential chances of replay attacks~\cite{hubaux2004security} \\
\hline
Leinm{\"u}ller \etal~\cite{leinmuller2006position,leinmuller2006improved,leinmuller2010decentralized} & OBU-C & Data-centric & Position cheating & Verify positions by: \ca discarding packets if the distance is farther\cite{leinmuller2006position}, \cb cooperatively exchange position information~\cite{leinmuller2006improved} \cc including additional checking~\cite{leinmuller2010decentralized} & Limited detection capabilities \\
\hline
Studer \etal~\cite{studer2007efficient} & OBU-L  & Data-centric & GPS spoofing & Estimate current position based on previous calculations & May cause approximation error \\
\hline
Zaidi \etal~\cite{zaidi2016host} & OBU-L & Data-centric & False data injection, Sybil attack & Perform statistical analysis to analyze traffic flow & Limited application scenario, Stealthy attacker may remain undetected \\
\hline
 Rawat \etal~\cite{rawat2011securing} & OBU-L & Entity-centric & False data injection & Predict vehicle  behavior using Bayesian logic & Potential classification errors, hard to obtain parameters \\
 \hline
 Kerrache \etal~\cite{kerrache2016t} & OBU-C, RSU & Entity and Data-centric & False data injection & Obverve vehicle behavior by local and global trust scores & May not work well if the adversarial behavior changes dynamically  \\
 \hline
 Raya \etal~\cite{raya2007eviction}, Moore \etal~\cite{motorway_attacker} & OBU-C & Entity-centric & False data injection & Find deviations from normal/average behavior  & Prone to Sybil attacks, potential false positives \\
 \hline
Zhuo \etal~\cite{zhuo2009removal} & OBU-C, RSU & Entity-centric & False data injection & Remove misbehaving
insiders from local and global analysis & Vulnerable to Sybil attacks \\
\hline
\end{tabular}
\end{footnotesize}
\end{table*}

\section{DoS and Sybil Attack Detection} \label{sec:dos_n_sybil}

In this section we first present solutions to detect DoS attacks (Section~\ref{sec:dos_detc}) and then describe various approaches proposed in literature for Sybil attack detection (Section~\ref{sec:sybil_dect}).

\subsection{DoS Detection/Mitigation} \label{sec:dos_detc}

Since DoS attacks~\cite{razzaque2013security} can be implemented at varying layers, researchers proposed different solutions to detect/mitigate the changes of attacks.
Jamming-based DoS attacks can be detected by behavioral mechanisms -- for instance, by analyzing the patterns in radio interference~\cite{hamieh2009detection} as well as by using statistical network traffic analysis and data mining methods~\cite{lyamin2018ai}.
The chances that (external) attackers to intrude/disrupt the system can also be reduced by using short-time private-public keys with a hash function~\cite{chuang2014team}. He \etal~\cite{he2012mitigating} proposed to use a pre-authentication process before signature verification to prevent DoS attacks against signature-based authentication (where attackers broadcast
forged messages with invalid signatures -- leading to unnecessary signature verifications). Researchers also proposed alternatives to digital signatures -- a new authentication method (called Tesla++)~\cite{studer2009flexible} that reduces the memory requirement at the receiver for authentication and can be used to limit the chances of resource (\eg memory) exhaustion. A downside of these protocols is a high delay between message arrival and message authentication.

Given the fact that the routing in V2X is predictable and standardized, network layer DoS attacks such as packet dropping
can be detected by watchdog mechanisms~\cite{watchdog_icc10} where each vehicle uses the idea of neighbor trust level (determined as the ratio of packets sent to the neighbor and the packets are forwarded by the neighbor). Packets may not be forwarded due to a collision and/or an attack. If a vehicle is repeatedly dropping packets (until a tolerance threshold is exceeded), the vehicle is considered as malicious -- although the evaluation results show that it is difficult to find a global threshold (\eg for deciding when misbehavior should be detected). Packet dropping/duplication can be prevented by clustering based monitoring~\cite{daeinabi2013detection} where a set of vehicles in a cluster (called verifiers) monitor the behavior of a (newly joined) vehicle. Vehicles that acted maliciously are blocked by certificate authority (CA) and are informed other vehicles.

There exist mechanisms~\cite{6799761} to detect flooding-based DoS attacks by observing channel access patterns -- for instance, by generating an adaptive threshold (that represents the maximum rate of messages any vehicle can send with respect to other vehicles). This approach may not be scalable for generic use-cases since the scheme is designed for vehicles communicating with a single RSU. Similar infrastructure-assisted mechanisms such as those proposed by Verma \etal~\cite{verma2013prevention,verma2015bloom} can prevent DoS attacks by: \ca monitoring V2X messages (that checks the number of outstanding packets with a  predetermined threshold within a certain window of time);
or \cb by using a message marking policy
where packets are marked by the edge routers (say RSUs) and if the sender IPs are found malicious, an alarm is sent to other vehicles. Recent work~\cite{biswas2012ddos} proposed to randomize the RSU packet transmission schedule and a modification of the congestion control schemes to mitigate packet flooding-based DoS/DDoS attacks. Message flooding can also be detected by trust-based mechanisms~\cite{hasrouny2017security,kerrache2017tfdd}. Hasrouny \etal~\cite{hasrouny2017security} propose to calculate trust values of the vehicles that can limit the number of accepted received messages from neighbors -- if a certain threshold is exceeded (which will be the case in DoS attack), a report is sent to the trusted entity, say misbehavior authority (MA)~\cite{vsc_pki}, to deactivate the attacker. The TFDD framework~\cite{kerrache2017tfdd} can detect DoS and DDoS attacks in a distributed manner by trust establishment between vehicles. Each vehicle maintains local and global parameters (\eg neighbor id, various message counters, trust score)  in order to include/exclude neighbours from a local or global black-list. A globally blacklisted vehicle can be suspended from network operations by the trusted authority~\cite{zhang2008highly}. The proposed mechanism may not work for  an intelligent, stealthy attacker whose (malicious) behavior may not remain stable throughout time.

\subsection{Detecting Sybil Attacks} \label{sec:sybil_dect}

Researchers proposed to detect Sybil attacks in V2X networks that can work either \ci  without any infrastructural support~\cite{grover2011sybil,hao2011cooperative,Golle:2004,yu2013detecting, guette2007sybil,ruj2011data} (Section~\ref{sec:sybil_no_rsu}) or \cii with assistance from infrastructure (\eg RSU, PKI, trusted authority)~\cite{xiao2006detection,sowattana2017distributed,chang2012footprint,zhou2011p2dap,chen2009robust,park2009defense,hamed2018sybil,lee2013dtsa,rahbari2011efficient,feng2017method,Singh:2017:RUA} (Section~\ref{sec:sybil_w_rsu}).

\subsubsection{Infrastructure-less Sybil Detection} \label{sec:sybil_no_rsu}

Grover \etal~\cite{grover2011sybil} suggest that the fake identities of the attacker must always be in the same vicinity  (for better control over malicious nodes) and proposed a detection by comparing the tables of several neighboring vehicles over time. This scheme does not protect against Sybil attacks that have a short duration. The communication overhead and detection latency is high, and certain scenarios (\eg traffic jams) may increase false positives or detection latency. Hao \etal~\cite{hao2011cooperative}  proposed a cooperative protocol that utilizes group signature (to preserve privacy) and correlation of mobility traces. The key idea is that vehicles around a possible attacker inform others by broadcasting warning messages with their partial signatures -- a complete signature can be derived (and hence the attacker is identified) when the number of vehicles that report anomalies reaches a threshold. The protocol is not verified for the case of multiple Sybil attackers.

A model-based approach, based on position verification, is proposed by Golle \etal~\cite{Golle:2004} where each node contains a model of the network and checks the validity of the received data using local sensors (\ie camera, infrared and radars). Data collected from the sensors can be used to distinguish between nodes. Inconsistencies can then be detected (\ie in case of Sybil attacks), based on the proposed heuristic mechanism, by comparing the received data with the model.  For instance, using a camera reading and exchanging data via a light spectrum a vehicle can verify whether a claimed position is true. Thus one can determine the real existence of the vehicle. However it is generally hard to obtain a generic model of the V2X network due to the dynamic nature and the proposed method is designed by considering high density road conditions only (\ie may not perform well for low density situations or roads where vehicle density varies over time).

Researchers also proposed the identification of falsified positions by exploiting channel properties, for instance, by analyzing its signal strength distribution~\cite{yu2013detecting, guette2007sybil,xiao2006detection} or by observing RSSI (received signal strength indicator) measurements~\cite{yao2019multi}.
A Sybil detection approach~\cite{guette2007sybil} analyzes physical layer properties under the assumption that antennas, gains and transmission powers are fixed and known to all the vehicles in the network. The authors use received signal strength to determine the approximate distance to the sender and further verify the transmitted GPS position. A similar idea is also used~\cite{ruj2011data} to verify locations by finding the co-relation between location, time and transmission duration (for both beacons and event messages). A post-event validation approach verifies specific event messages (by analyzing messages from other vehicles) and also pseudonym change mechanism is applied once a claimed event is detected as being malicious (\eg that is detected by a lack of subsequent beacon messages from the same source). This scheme, however, can be exploited to revoke legitimate vehicles by an attacker with jamming capabilities (since they are based on physical-layer signal properties)~\cite{misbhv_survey_2019}. Prior work~\cite{guette2007sybil,ruj2011data} also do not account for GPS errors in the model.
A distributed  mechanism, Voiceprint~\cite{yao2019multi}, was proposed to detect Sybil attacks by analyzing RSSI (received signal strength indicator)-based measurements (\eg by performing a similarity measures between the RSSI of an attacker and its Sybil nodes over time). However Voiceprint may not detect attacks if an adversary uses more than one radio.

\subsubsection{RSU-assisted Sybil Detection} \label{sec:sybil_w_rsu}

There exist mechanisms~\cite{xiao2006detection,sowattana2017distributed,chang2012footprint,zhou2011p2dap,chen2009robust,park2009defense,hamed2018sybil,lee2013dtsa,rahbari2011efficient,feng2017method,Singh:2017:RUA} to use a centralized authority (\eg RSU) to detect Sybil nodes. In an earlier study Xiao \etal~\cite{xiao2006detection} verify claimed positions using signal strength metrics where vehicles are assigned three roles: \ci \textit{claimer} (a vehicle claims a position using a beacon), \cii  \textit{witness}, (a node receives a beacon and measures its proximity using the received signal strength that is then transmitted in subsequent beacons) and \ciii \textit{verifier} (the vehicle that collects signal strength measurements to estimate and verify the position of a vehicle). RSUs issue signatures of vehicles in their proximity at a specific time along with a driving direction. When a beacon message is received, the verifier waits for a period of time (to collect previous measurements of the claimer from witness) and calculate an estimated position of the claimer.
A similar idea is used in a consensus-based Sybil detection scheme~\cite{sowattana2017distributed}: each receiver validates the validity of received beacons by transmission range of those neighbors and generates a trust score using a voting scheme. The voting process itself, however, is vulnerable to the attack.

Researchers also proposed~\cite{park2009defense} to use message timestaps (\eg to find each vehicle's recent trajectory and time) for Sybil detection that do not require any PKI. Before sending any messages, a vehicle first  obtains a timestamp for the message from a nearby RSU. If a vehicle receives similar timestamp series from the same RSUs for a certain amount of time then that vehicle is considered as Sybil node. However two vehicles coming from opposite directions could be incorrectly marked as Sybil nodes since they will receive similar timestamps for a short time period. The P\textsuperscript{2}DAP framework~\cite{zhou2011p2dap} detects Sybil attacks by identifying vehicles with different pseudonyms and propose an inherent linking between pseudonyms based on hash functions. For instance, a message is considered malicious if the tuple, (time, location, event type), is signed by the same vehicle with different pseudonyms. The (semi-trusted) RSUs are responsible for checking messages the linking and reports it to the central authority (that can resolve pseudonymity). However the ability to link arbitrary pseudonyms may be a privacy issue. Time and spatial granularity as well as event types needs to be standardized and also the central authority requires the complete knowledge of the network. Similar architecture was used in recent work~\cite{hamed2018sybil}. It leverages the fact that two
vehicles may not pass multiple RSUs at the same time. The authors proposed to detect Sybil nodes by monitoring mobility patterns of vehicles. This scheme requires a global view of the network and also the estimation of a proper detection threshold is not straightforward (may result in false positive/negative errors).

Chen \etal~\cite{chen2009robust} identify that Sybil nodes that originate from the same vehicle will always have similar trajectories over time. With this observation a new protocol uses special signatures (obtained from RSUs) that can be used to build a reference trajectory. Identities with identical recent trajectories are considered to be the same vehicle thus minimizing the effect of Sybil attacks. However there exists practical limitations: \ca bandwidth overhead for signature exchanges; \cb potential chances of DoS attacks -- since each request requires a much larger response (\eg the signatures) and \cc privacy violations -- vehicles need to reveal their position traces (\ie set of signatures) to verify themselves as non-Sybil nodes. The  ideas were also extended in the privacy preserving Footprint framework~\cite{chang2012footprint} where cryptographically protected  trajectories (consisting of special signatures) are requested by the vehicle from the RSUs. The trajectories for every message are used as an authentication mechanism that allows a vehicle to compute the Sybil nodes (\eg when all trajectories that are suspiciously similar are coming from a same vehicle). Footprint protects vehicle privacy (\eg from long-term tracking) since signatures of RSUs are time-dependent (and unpredictable).

Another privacy-preserving protocol -- DTSA\cite{lee2013dtsa} uses session key-based certificates where each vehicle's (unique) ID is registered to a global server. The vehicles then generate anonymous IDs that are validated by a local server (and a local certificate is issued). Any receiving vehicle can verify the message by comparing the other vehicle's true identity with the local server. This scheme may reduce network throughput since certification exchanges require a large amount of overhead data. Similar ideas exist in earlier work~\cite{rahbari2011efficient} that utilizes PKI and a local CA to detect Sybil attacks by comparing the reply message received from the RSU. More recent framework EBRS (event based reputation system)~\cite{feng2017method} proposed to use short-term public key and pseudonyms that needs to be validated by a trusted authority (\eg by using RSUs). While EBRS can detect attacks from multiple sources, this framework may collapse if the RSUs and/or OBUs are compromised. Researchers also proposed to use anonymous credentials (\ie a specific certificates) and a cryptographically protected usage restriction of the credentials -- since the sender is allowed to use only one credential per time  Sybil nodes then can be detected~\cite{chen2011threshold,Singh:2017:RUA}. However the performance overheads of these approaches are high compared to the ECDSA algorithm proposed in the standards~\cite{ieee_wave,etsi_itsc}.





\section{Integrity Checking} \label{sec:integrity_check}

The integrity of V2X communication can be verified from different contexts such as: \ci validating events (Section~\ref{sec:int_event_val}), \cii checking message integrity (Section~\ref{sec:int_message_int}), \ciii location verification (Section~\ref{sec:int_location_verif}) and \civ reputation analysis (Section~\ref{sec:int_rep_analysis}) as we discuss in the following.

\subsection{Event Validation} \label{sec:int_event_val}

Kim \etal~\cite{kim2010vanet} propose a message filtering mechanism that combines parameters of messages into a single entity called the `certainty of event' (CoE) curve. CoE represents the confidence level of a received message and is calculated by combining the data from various sources such as local sensors and RSUs and by using consensus mechanisms (\eg messages from other vehicles and validation by infrastructure, if available). Message validity is defined using a threshold curve and false positives for events can be reduced when more evidence is obtained over time. While the mechanism is applied to the the emergency electronic break light application (\eg that enables broadcasting self-generated emergency brake event information to nearby vehicles), it unclear how this scheme behaves for generic V2X applications (say for multiple lanes and urban settings where there may be some uncertainty about the vehicle paths) since it requires specific locations for the events. Besides, such CoE-based mechanisms could be vulnerable to Sybil attacks depending on how the information from other sources are captured.

Researchers also proposed to determine the correctness of event reports through voting~\cite{cao2008proof}  -- the key idea is develop an efficient way to collect signatures from a sufficient number of witnesses without adding too much (bandwidth) overheads on the wireless channel. If insufficient signatures are received, events may be missed completely (\ie may cause false negative errors). A similar idea is also used by Hsiao \etal~\cite{hsiao2011efficient} where the senders collect a number of witnesses for each possible event. However this model enforces a specific message format and there is no deflation protection, \ie the attacker can reduce the amount of signatures attached to the message and/or can hide events. A consensus-based mechanism is proposed~\cite{petit2011spoofed} where each vehicle collects reports  about the same event from neighboring vehicles until a certain threshold of supporting reports is passed (after which the message is considered to be trustworthy). The proposed method allows the system to reach a decision within a bounded waiting time and thus suitable for time/safety-critical applications (\eg the decision whether to trust the warning about traffic accident that must be made early so that the vehicle can slow down or change lanes accordingly). Similar to the most consensus-based mechanisms, this approach also suffers from potential Sybil attacks.

The idea of post-event detection~\cite{ghosh2010detecting} can also be used for event validation: for instance, in post-crash notification (PCN) applications, once a PCN message is sent drivers adapt their behavior to avoid crash site and this information (\eg drivers behavior) can be used to identify whether the event was valid or not. The key idea is to use a technique (called root cause analysis) to detect which part of the event message was false (\eg upon receiving a PCN alert, the vehicle analyzes the sender's behaviors for a while and compares the actual trajectory and the expected trajectory). Such detection approaches suffer if the driver behavior models are fragile -- although this may not be a limiting factor for autonomous driving where valid driver behavior will be more well-defined.

\subsection{Behavioral Analysis and Message Integrity Checking} \label{sec:int_message_int}

The VEBAS (vehicle behavior analysis and evaluation scheme) protocol~\cite{schmidt2008vehicle} allows the detection of unusual vehicle behavior by analyzing all messages received from neighboring vehicles. VEBAS uses a a trust-based mechanism, \eg once a vehicle has collected information about surrounding vehicles, it will then broadcast the results (\eg trust-scores) within the single hop neighborhood. This checking mechanism uses a combination of behavioral mechanisms (\eg frequency of sending beacons) and physical parameters such as velocity and acceleration to determine the authenticity of a message. However VEBAS could be vulnerable since there is no mechanism the verify the correctness of the messages received from the neighbours.

The MisDis protocol~\cite{yang2013misdis} ensures
accountability of vehicle behaviour by recording all the (sent/received) messages for each vehicle peer in a secure log. Any vehicle can request the secure log of another vehicle and independently determine deviation from expected behavior. This protocol, however, requires strong identification and authentication mechanisms and there is no discussion about how the vehicle privacy is preserved. Also authors do not provide any performance evaluation of the proposed method. Lo \etal~\cite{lo2007illusion} propose a plausibility validation network (PVN) to protect the V2X applications from false data injection attacks (called illusion attacks) where attacker can indirectly manipulate messages (\eg through sensor manipulation). The idea is to use a rule database (\eg a database of rules specifies whether a given information should be considered valid or not) and a checking module that checks the plausibility of the received messages.
Each message is evaluated with respect to its type (accident report, generic road condition) and the corresponding predefined rule set is retrieved from the rule database to check the value of the message element fields (\eg timestamp, velocity).
 For instance, the plausibility of the timestamp field is checked by determining the minimum and maximum bounds \eg the received timestamp must be earlier than the receiver’s current timestamp $t_c$ and later than the difference between the $t_c$ and the validity period of the message.
 A limitation of this approach is that since the rule database is shared, a malicious vehicle can generate valid messages to avoid detection.

\subsection{Location and GPS Signal Verification} \label{sec:int_location_verif}

Researchers used different techniques to predict the position and behavior of vehicles (\eg whether they follow an expected pattern) in order to identify malicious vehicles. One idea is to verify node positions using two \textit{verifiers}~\cite{vora2006secure}: acceptors (distributed over the region) and rejecters (placed around acceptors in circular fashion) -- say for a given region, by using multiple RSUs (rejectors) surrounding one (center) RSU (acceptor). If the message is first received by the acceptors, then they will verify that the vehicle is within the region. However a malicious vehicle can spoof its location when it resides within the region since the protocol does not verify the exact location of the nodes. Yan \etal~\cite{yan2008providing} proposed to use on-board radar to detect the physical presence of vehicles (\eg for applications such as a congestion alert system). The vehicles compare radar information (\eg which vehicles are in proximity) with the GPS information received from other vehicles to isolate malicious nodes. The mechanism can prevent some variants of Sybil attacks, \eg by calculating the similarity of radar information, reports from neighbours and oncoming traffic reports. There exist mechanisms~\cite{stubing2010verifying,jaeger2012novel} to verify transmitted CAMs by analyzing the sequence of messages (\eg to find the trajectory of each vehicle). By tracking a vehicle (say by using a Kalman filter\footnote{Kalman filters can accurately predict the movement even under the influence of errors -- for instance, they can be used to correct errors in GPS measurements~\cite{misbhv_survey_2019}.}), the receiver can verify the location contained within each CAM. The idea is extended~\cite{stubing2011two} to applications where the accuracy of the Kalman filter  is poor (\eg for special maneuvers or lane changes scenarios). A signature-based scheme~\cite{bissmeyer2010intrusion} based on a plausibility checking is proposed where each vehicle is modelled as differently sized (and nested) rectangles -- intersecting rectangles that belong to different vehicles indicate false position information.  Since the readings from positioning systems (\ie GPS) could be inaccurate, the probability of intersections is calculated by intrusion certainty (based on the number of observed intersections) and trust values (\eg using minimum-distance-moved concept~\cite{schmidt2008vehicle} where any neighboring vehicle $V_j$ who is further than a given vehicle's transmission range is considered more trustworthy). When $V_j$ intersects with another neighbor and the difference between trust levels of both vehicles is higher than a predefined threshold then the less trustworthy vehicle is considered to be malicious. While this method can detect false positions despite GPS errors, an attacker with larger transmission ranges (compared to other vehicles) can bypass this mechanism.

Vehicle positions can be verified by physical properties such as Doppler speed measurements of the received signal~\cite{8228654}. The idea is to use the angle of arrival (AoA) and Doppler speed measurements. When this information is combined with the position information included in the message, the estimation error (calculated using an extended Kalman filter based approach) should not diverge unless the vehicle misbehaves by transmitting false location information. Another approach to verify vehicle position is \textit{distance bounding}~\cite{brands1993distance} -- a technique to estimate distance using physical characteristics such the speed of light. Since light travels at a finite speed, an entity (\eg RSU or other vehicle) can measure the (round-trip) time to receive a message and determine an upper bound on the vehicle distance. By using distance bounding mechanisms Hubaux \etal~\cite{hubaux2004security} show that RSUs can verify a vehicle's location when: \ci  three RSUs are positioned to form a triangle (for a two dimensional plane) or \cii four RSUs form a triangular pyramid (for a three dimensional plane). In a similar direction, researchers proposed a data-centric mechanism to verify false position information using timestamps~\cite{ruj2011data}. For example, when location information $L_i$ (timestamped at $t_i$) is received by a vehicle (located at $L_j$) at time $t_j > t_i$, the receiver can verify the correctness of this information using the locations, speed of light and the difference between timestamps. A malicious vehicle cannot modify timestamps (say $t_i$) since the exact location between the attacker a receiver vehicle is unknown. When a false location is detected the receiver broadcasts this information to other vehicles (and perhaps to the CA via RSU).

An attacker can send delayed responses to each RSU~\cite{hubaux2004security} (\eg by using directed antennas), An alternative trust-based position verification approach is proposed
where a vehicle discards packets if the included position information is further than the predefined maximum acceptance range threshold~\cite{leinmuller2006position}. Since the recipient negatively weighs abnormal observations (\eg the sender's trust level is more affected by abnormal observations), after sending one bogus information packet a (malicious) vehicle is required to send correct information packets in order to regain its previous trust level. Similar ideas can be used by exchanging position beacons among neighbors~\cite{leinmuller2006improved}, \ie  beacons received from neighbors are checked against received neighbor tables by comparing the (claimed) positions for a particular node in the beacon and the table. These mechanisms can be improved by \ci ignoring further beacons when too many of them are sent from one area (\eg to limit the impact of potential Sybil attack) \cii map-based verification (\eg by assigning a plausibility value to the received beacons by comparing the location to the road map) and \ciii position claim overhearing (\eg for geo-routing scenarios by comparing different overheard packets and respective destinations can provide indications of a false position in the past)~\cite{leinmuller2010decentralized}. All of these checks, however, may not perform well individually~\cite{van2016enhanced}. 

There exist mechanisms to detect GPS spoofing by \textit{dead reckoning}, \eg where the current position is calculated by using a previously determined position and known (or estimated) speeds over elapsed time~\cite{studer2007efficient}. While this method can detect spoofed GPS information, the calculated position is only an approximation. For details of GPS spoofing countermeasures and recent proposals we refer the readers to further related work~\cite{bittl2015emerging, gnss_survey_2016}.

\subsection{Reputation Analysis and Revocation} \label{sec:int_rep_analysis}

Researchers have also proposed mechanisms such as statistical analysis and explicit voting to decide trustworthiness of the vehicles. Zaidi \etal~\cite{zaidi2016host} use statistical techniques  to predict and explain the trends in traffic flow and determine whether or not a sender is malicious. Each vehicle $V_i$ estimates its own flow parameter $F_i$ (that should be similar for vehicles located closely to $V_i$) by using a model (that uses vehicle density per and the average speed of other vehicles in its vicinity). Vehicles exchange their own flow parameters, density values, speed and location information. For each received message, vehicles compare the average of the received parameters to its own calculated parameters -- if the difference is lower than a predetermined threshold then the message is accepted; otherwise, the behavior of the sender is monitored (\ie only accept messages until it is enough to perform a statistical test). The malicious vehicle will then be reported to other vehicles and isolated from the network. A stealthy attacker (one who manipulates values gradually), however, may remain undetected. An approach using Bayesian logic has proposed to compute the `probability of maliciousness' of a vehicle for a time $t$, given some observation $O_t$~\cite{rawat2011securing}. The idea relies on Bayesian reasoning, \ie computing the probability of the vehicle being malicious given $O_t$ (\eg by applying Bayes' theorem). This scheme requires prior knowledge of the probability of reception of a particular message and the authors do not specify how these conditional probabilities can be obtained for generic V2X use-cases. The T-VNets framework~\cite{kerrache2016t} evaluates two trust parameters: \ca inter-vehicles trust (\eg by combining data-centric evaluation of messages received from each neighbor) and \cb RSUs-to-vehicles trust (built by collecting reports from vehicles about their neighbor's behaviors -- to build a quasi-global historical and regional trust value). The authors propose to periodically exchange global trust values  by adding the addition of new fields to the CAM messages. Besides, DENMs are used to dynamically calculate the trust for specific events (\eg road hazards) -- the events that have a lower trust value than a predefined threshold will not be broadcast by the vehicles. However the authors assume attackers always and persistently exhibit dishonest behavior throughout time and that may not be the case in practice.

Raya \etal~\cite{raya2007eviction} proposed LEAVE (local eviction of attackers by voting evaluators): an entropy-based measurement with k-means clustering to detect which neighbor differentiates from other neighbors (\eg a misbehaving vehicle) -- say if high velocity information received from a neighboring (malicious) vehicle is contradictory to messages from the majority of vehicles (\eg for a traffic jam situation) then the malicious vehicle will be detected. Vehicles exchange `accusations' about potential attackers and the malicious vehicle can be evicted temporarily (by revoking its certificate).  A core advantage of LEAVE is the reduced detection latency (since vehicle trust does not need to be built over time). A similar idea is also proposed by Moore \etal (called Stinger)~\cite{motorway_attacker} in which both the reporting as well as reported vehicles are temporarily prohibited from sending messages. Both LEAVE and Stinger protocols require an honest majority -- if there exists too many compromised neighbors then they could present malicious behaviors as normal (\eg vulnerable to Sybil attacks). Zhuo \etal~\cite{zhuo2009removal} proposed a cooperative local and global eviction mechanism:  SLEP (a so-called suicide-based eviction mechanism that is designed to discourage false accusations) and PRP (that uses trust level of each accuser to decide on permanent revocation) respectively, to remove misbehaving vehicles. The basic idea is that if a vehicle can detect bogus messages (say by comparing on-board sensor information about the event), it will broadcast a message accusing the potential attacker vehicle (and the neighboring vehicles will then ignore the messages from accused vehicle). In  contrast to other work~\cite{raya2007eviction} a vehicle can use pseudonyms (\ie to protect privacy) and can re-join the network after a successful accusation. Limitations of exiting revocation schemes include~\cite{liu2010limits}: \ca they assume   a local honest majority and if an attacker manages to create a local majority (that is the case of Sybil attacks) then it is possible to create false accusations (and falsely remove honest vehicles from the network) and \cb when pseudonyms are used (\ie to protect user privacy) an attacker can use multiple pseudonyms in parallel to create a local majority. For voting-based schemes researchers therefore suggest \textit{not} to use multiple pseudonyms in parallel (\ie they should be prevented by the underlying pseudonym mechanism)~\cite{liu2010limits}.






\section{Discussion} \label{sec:discussion}

In this section we briefly highlight open issues both for IEEE 802.11p-based (Section~\ref{sec:dsrc_openissue}) and LTE-based V2X communications (Section~\ref{sec:cv2x_sec}). We then review security issues related to low-level protocols running within a vehicle (Section~\ref{sec:can_sec}).








\subsection{Open Issues and Design Considerations} \label{sec:dsrc_openissue}

As we mentioned in Section~\ref{sec:attacks_def}, the robustness of V2X technology (due to predefined packet authentication and use of timestamps) mitigates the severity of spoofing attacks (\eg replay or man-in-the-middle attacks). While the use of digital signatures and PKIs have been widely studied and standardized for V2X communication, there is a gap between existing academic research and large scale practical testing of PKI for V2X applications. It requires further investigation and experiments to discover (and resolve) potential issues including ambiguous specifications in standards, equipment interoperability from different vendors and scalability~\cite{den2018security}. There exists a trade-off between different aspects such as false positive rate, CRL size, complexity, RSU availability. In addition, most of the existing
V2X security solutions are known for their
high computation and delay overheads (see
Sections~\ref{sec:dos_n_sybil} and~\ref{sec:integrity_check}). We also observe that \ca experimental
evaluation and benchmarking of these security solutions
have only been conducted under limited operating conditions and \cb there exists a lack of evaluation, comparison and feasibility study for the existing methods. An important problem in V2X security solutions is that of configurations. For instance, after what threshold should a message/event/activity should be considered as malicious? This is itself an important research challenge since high false positive/negative rates can easily destabilize any security technique.

There  are still remain  open questions regarding CRL distribution and pseudonym change strategies.
Modern traffic
analysis techniques can also examine traffic patterns and extract location information~\cite{yeh2016security}. However, in order for an attacker to track a vehicle based on BSM/CAM, an attacker needs to follow the transmitter vehicle to be in close proximity.  Pseudonyms may not be sufficient to prevent location tracking since an attacker can infer complete travel paths by combining pseudonyms and location information~\cite{wiedersheim2010privacy}.

While most of the related work focuses on detecting misbehaving vehicles, designing efficient response mechanisms still an open issue -- this is crucial especially for DoS/DDoS attacks where it is almost impossible to respond to the attack.
Often solutions proposed in literature assume RSUs are fully trusted \cite{6799761,verma2013prevention,verma2015bloom,he2012mitigating,studer2009flexible,biswas2012ddos,xiao2006detection,sowattana2017distributed,chang2012footprint,zhou2011p2dap,chen2009robust,park2009defense,hamed2018sybil,lee2013dtsa,rahbari2011efficient,feng2017method,Singh:2017:RUA,vora2006secure}. This may not always be the case in practice since RSUs are deployed roadside and may be susceptible to physical attacks (\eg sensor tampering,  differential power analysis). Therefore, there is a requirement for layered defense mechanisms that consider potentially vulnerable RSUs. Another (perhaps less technical) challenge is that of widespread implementation (\eg installation and maintenance) of V2X-compatible infrastructure and vehicular fleets -- the costs for RSUs and the PKI could be one of the biggest obstacles for full V2X deployment~\cite{machardy2018v2x}.

\subsection{Security Issues for LTE-V2X} \label{sec:cv2x_sec}


3GPP recognizes the need for user authentication (\eg only authorized entities should be able to transmit data) and suggests the processing of messages whose data origin has been verified by the vehicle~\cite{3gpp_sec1}. 
3GPP also states that vehicle identity should not be long-term trackable or identifiable from its transmissions. To achieve this, permanent identities of the vehicles need to be properly protected (and also exposure minimized say by using pseudonyms). This is important since fake base-stations can force legitimate vehicles to share their IMSI (international mobile subscriber identity) and/or location information~\cite{norrman2016protecting} and thus could be vulnerable to multiple classes of attacks (\eg Sybil and data injection).

While the use of temporary pseudonymous certificates (for vehicle authentication) provide a measure of privacy for DSRC/C-ITS, the association with a subscriber ID in LTE-V2X pose a threat of potential compromise of vehicle privacy, especially considering cellular network operators~\cite{machardy2018v2x}. Although 3GPP Release 14 (TS33.185)~\cite{3gpp_rel14_v2x_sec} specifies security requirements for LTE-V2X, the specifications do not yet impose any privacy mechanisms for the LTE-V2X PC5 (leaves this to the regional regulators and operators).
While 3GPP suggests changing and randomizing the layer 2 ID and IP address of the source (along with changing the application layer ID),  there is no additional protection for the Uu apart from what current LTE networks support.


There also exist unique issues of LTE-V2X (\eg maliciously mimic and/or control behavior of the base stations) due to centralized control in Uu-based LTE-V2X and PC5 mode 3. For example, if an attacker gains control of base stations, the attacker can \ca fully control scheduling of Uu-links as well as sidelinks (PC5 mode 3), \cb allocate collided resources to vehicles to degrade the communication performance, \cc provide a wrong network configuration to the vehicles and \cd obtain location information. LTE-V2X PC5 mode 4 and DSRC/C-ITS, however, are not vulnerable to such issues as they operate in a fully distributed manner.

\subsection{Threats to Intra-vehicle Components and Countermeasures} \label{sec:can_sec}

Modern vehicles are equipped with a swarm of sensors, camera, radar, LiDAR that can be tampered by the adversary.
 Possible attack surfaces
(\ie from where the attack could originate) include~\cite{petit2015potential}: \ca vehicle sensors, \eg acoustic sensors,  odometric sensors (such as wheel encoders, accelerometers, gyroscope), radar, LiDAR and vision systems (used for object detection), GPS modules (used for localization and positioning) and \cb in-vehicle user devices that can be connected to the infotainment system
via Bluetooth/WiFi/USB. Although intra-vehicle (\eg on-board) attacks are not directly related to communication/network security, such attacks could prevent the vehicle from operating normally and destabilize V2X communication networks. For instance, intra-vehicle attacks such as \textit{side-channel attacks} can lead the attacker to infer the secret information (\eg cryptographic keys)~\cite{jain2018probing}
or \textit{DoS attacks} (\eg that disables the steering/braking/LiDAR/camera system in an advanced driver assistance systems and automated driving systems) could disrupt the normal operation of the vehicle and/or pose threat to human safety~\cite{andrews2017guidelines, miller2014survey, petit2015potential, raya2007securing}. Recent work on the security of controller area networks (CANs)~\cite{hoppe2008security, checkoway2011comprehensive, kleberger2011security} -- the in-vehicle communication bus used in some vehicles -- has shown that they are vulnerable to such attacks. Given the fact that the vehicles in the V2X network can be connected to untrusted mediums such as Internet (\eg by RSUs), and therefore, the sub-systems, ECUs/OBUs could be remotely compromised/controllable~\cite{miller2014survey, woo2015practical, ishtiaq2010security}. One way to address this problem is to use a central gateway that enables secure and reliable
communications among a vehicle's electronic systems~\cite{gateway1, gateway2}.

One of the major concerns for securing in-vehicle architecture is to protect the hardware and applications running inside ECUs. Researchers has proposed different techniques such as: \ci use of hardware security modules (HSMs) for secure boot, processing and storage~\cite{apvrille2010secure}; \cii various isolation mechanisms (\eg by using virtualization, container, microkernel, \etc)~\cite{groll2009oversee,stumpf2009security}; \ciii hardware/software architecture for over-the-air (OTA) updates~\cite{idrees2011secure}; \civ statistical analysis of ECU firmware images by reverse engineering to detect misbehaving ECUs~\cite{contag2017they}, \etc~to name but a few. Despite the isolation mechanisms, vehicles may still remain insecure due to implementation bugs  and/or poor isolation policies. Besides, verification of policies/implementations requires enormous effort for such complex automotive platforms.

Given the vulnerabilities of the CAN bus~\cite{checkoway2011comprehensive,palanca2017stealth,murvay2017attacks}, a number of mechanisms have been proposed: \ci encrypting CAN messages and hiding system states to protect against selective DoS attacks~\cite{glas2012signal}; \cii use of authentication schemes (for both ECUs and CAN messages) to ensure their integrity~\cite{wolf2006secure,nilsson2009defense,van2011canauth,groza2012libra,lin2015security,radu2016leia,zou2017study}; \ciii use of asymmetric cryptography and certificates to authenticate ECUs and share symmetric keys~\cite{mundhenk2017security}. Researchers have also studied the use of behavioral-based intrusion detection systems (IDS) for in-vehicle networks. However, building such IDS for in-vehicle networks is challenging due to the large number and heterogeneity of ECUs as well as due to limited information exposed by CAN messages (since they are specific to manufacturers and/or vehicle model)~\cite{den2018security}. While there exist IDSes for in-vehicle networks (\eg by utilizing  message frequency~\cite{hoppe2009applying,song2016intrusion,taylor2015frequency}, entropy~\cite{marchetti2016evaluation}, clock skew~\cite{cho2016fingerprinting}, observing cyber-physical contexts~\cite{wasicek2017context}) these systems may not be able to detect attacks involving sporadic/irregular CAN messages. Researchers also proposed to replace  the CAN technology and use other alternatives such as Ethernet~\cite{bello2011case,tuohy2015intra,hank2012automotive}. While earlier work focus on improving bandwidth and reducing latency/error rates, the impact of Ethernet on vehicle security is not thoroughly investigated and require further research. We also highlight that CAN will most likely remain as the most common in-vehicle networking technology over the next decade~\cite{hank2012automotive}. Replacing CAN will not solve all security/privacy issues and security measures (such as IDSes) built on top of CAN will remain applicable even when CAN has been replaced~\cite{den2018security}.



\section{V2X Security Projects and Related Work}

The vehicular communication sector has been widely studied. In this section we first provide
a list of main academic and industrial research projects actively working on various aspects
 of V2X security (Section~\ref{sec:projects}). We then summarize related surveys that discuss security and privacy issues in the context of V2X applications (Section~\ref{sec:rel_work}).

\begin{table*}[h]
\centering
\caption{Qualitative Comparison of the Major V2X Security Projects in Europe and United States}
\label{tab:v2x_sec_projects}
\begin{footnotesize}
\begin{tabular}{P{2.5cm}||P{2.1cm}|P{2.1cm}|P{2.1cm}|P{2.1cm}|P{2.1cm}|P{2.1cm}}
 & EVITA & sim\textsuperscript{TD} & OVERSEE & PRESERVE & ISE & CAMP-VSC6 \\
\hline \hline
Project focus\textsuperscript{a} & OBS & CNS & OBS & OBS and CNS & CNS & CNS  \\
\hline
Objective & On-board intrusion detection/prevention & Secure V2X communications & Secure and standardized communication/application platform  & Close-to-market security/privacy solution for inter-and-intra-vehicle networks & Privacy-preserving message authentication & Security credential management and misbivevior detection \\
\hline
Evaluation approach &  Proof-of-concept implementation & Field trial, simulations, conceptual\textsuperscript{b} & Proof-of-concept implementation & Proof-of-concept implementation, simulations & Proof-of-concept implementation & Conceptual\textsuperscript{b}, prototype development (ongoing)\\
\hline
Reuse of existing projects & No & No & Yes\textsuperscript{c} & Yes\textsuperscript{d} & No & No\\
\hline
Use of PKI & N/A & Yes & N/A & Yes & Yes & Yes\\
\hline
Initiative & European Union & Germany & European Union & European Union\textsuperscript{e} & France & United States\\
\hline
Status & Completed  (2008-2011)  & Completed (2008-2013) & Completed  (2010-2012) & Completed  (2011-2015) & Completed  (2014-2017) & Ongoing (2016-present)\\
\hline
\end{tabular}
\begin{flushleft}
\hspace{2em}\textsuperscript{a}Scope of the security analysis -- OBS: Intra-vehicle (on-board) security, CNS: Inter-vehicle (communication/networking) security. \\

\hspace{2em}\textsuperscript{b}Conceptual/analytical/architectural demonstration of the system components -- not implemented/verified/tested on real systems. \\

\hspace{2em}\textsuperscript{c}Used the concept/implementation of hardware-based security module from EVITA project.\\

\hspace{2em}\textsuperscript{d}Reused various components from multiple past projects (\eg EVITA,
sim\textsuperscript{TD}, \etc) \\

\hspace{2em}\textsuperscript{e}CAMP consortium was also a supporting partner of this project.
\end{flushleft}

\end{footnotesize}
\end{table*}

\subsection{V2X Security Projects} \label{sec:projects}


During the last decade there has been the rise of several research and development projects focusing on securing V2X communications with a view to design, analyze and test suitable security mechanisms. Table \ref{tab:v2x_sec_projects} summarizes a comparative study of the various V2X security projects in the United States and Europe.

The EVITA project\footnote{https://www.evita-project.org/} aims to  develop a secure internal on-board architecture and on-board communications protocols to prevent and/or detect illegal tampering. It also considered legal requirements of on-board networks with respect to privacy, data protection and liability issues. The sim\textsuperscript{TD} project\footnote{http://www.simtd.de} investigated the contribution of secure V2X systems for improving traffic safety and mobility using real-world field tests. The project developed different concepts, protocols, cryptographic procedures and privacy preserving mechanisms for the V2X field trials. The OVERSEE project\footnote{https://www.oversee-project.com/}  proposed a secure, open in-vehicle platform, for the execution of OEM and non-OEM applications.
This project aims to develop protected runtime environments (for the simultaneous and secure executions) by providing isolation between independent applications. It also proposes to provide a secure interface from the outside world to the internal network of the vehicle. The various  security and privacy aspects (\eg performance, scalability, and deployability) of future V2X systems is addressed in the PRESERVE project\footnote{https://www.preserve-project.eu/}. PRESERVE was one of the main European projects that experimented with multiple V2X security/privacy solutions and the design and implementation efforts were proposed to the standardization bodies. The ISE project\footnote{https://www.irt-systemx.fr/en/project/ise/} aims to design and implement a PKI system that is compatible with ETSI standard~\cite{etsi_pki}. The CAMP (crash avoidance metrics partnership) VSC6 (vehicle safety communications 6) consortium proposed the detection of misbehavior (\eg inadvertent transmission of incorrect data) in the V2X network both in local and network-level (\eg using in-vehicle
algorithms and processing as well as using a security credential management
system (SCMS)~\cite{vsc_pki}). This research prototype is now one of the leading candidates to support the establishment of PKI-based V2X security solution in the United States.


\subsection{Related Surveys} \label{sec:rel_work}


A number of surveys have been published on various aspects of the vehicular communications in the last decade. In prior work Saini \etal~\cite{saini2015close} provide a meta-survey of existing research for generic VANET (vehicular ad hoc network) domain. There also exists early research discussing application/platforms~\cite{simonot2009vehicle} and communication technologies~\cite{coppola2016connected}. However vehicular security and privacy aspects are not well studied. Prior work~\cite{kleberger2011security,petit2015potential,parkinson2017cyber} briefly reviews the security and privacy issues of in-vehicle protocols (\eg CAN) -- although communication aspects and standardization activities are not discussed.

Security and privacy issues in conventional vehicular networks have also been largely studied and there exist multiple surveys~\cite{verific_survey,azees2016comprehensive,vanet_iov_survey,local_coop_survey18,lu2018survey,hamida2015security,hasrouny2017vanet,machardy2018v2x,den2018security,kerrache2016trust,kamel2018feasibility,misbhv_survey_2019} that discuss several aspects (\eg functional requirements, protocols, vulnerabilities, \etc). In an early study~\cite{verific_survey} researchers survey various misbehavior (both faulty and malicious) detection approaches and countermeasures against spreading malicious data in the vehicular networks. However this work is primary focus on false data injection attacks and does not cover the broader scope of the field. Azees \etal~\cite{azees2016comprehensive} study VANETs as a special case of mobile ad-hoc network -- a common view in the past --  and does not cover the class of attacks against safety-critical systems (\eg false data injection) as is the case for modern V2X applications. Recent work~\cite{vanet_iov_survey} also surveys detection mechanisms for various classes of vehicular communication attacks (\eg DoS and network layer attacks). However the above work primarily focuses on routing-oriented attacks and defence mechanisms. Arshad \etal~\cite{local_coop_survey18} summarize the false information detection techniques for generic VANETs. Lu \etal~\cite{lu2018survey} survey  anonymous authentication schemes and Hamida \etal~\cite{hamida2015security} study challenges related to the
secure and safe V2X applications for ETSI C-ITS standard -- although their primary focus is on cryptographic countermeasures. In contrast our survey aims to provide a general overview of the security aspects of the modern V2X (\eg DSRC/C-ITS as well as C-V2X) platforms/applications.

 Recent surveys by Hasrouny \etal~\cite{hasrouny2017vanet} and MacHardy \etal~\cite{machardy2018v2x} provide a broad overview of the V2X communication including different radio access technologies, standardization efforts, attack techniques as well as security issues. Le \etal~\cite{den2018security} also studied the security and privacy requirements both, from intra- as well as inter-vehicle perspective. However, due to the very broad scope, all of the aforementioned work does not provide sufficient details on detection mechanisms.  A survey of the existing trust models for VANETs has also been carried out~\cite{kerrache2016trust}.
Kamel \etal~\cite{kamel2018feasibility} study multiple misbehavior detection methods and
then discuss their feasibility with respect to current standards, hardware/software requirements as well as with law compliance. Recent work~\cite{misbhv_survey_2019} studies misbehavior detection mechanisms for V2X applications -- although authors mainly focus on the DSRC/C-ITS context, and unlike us, they do not provide details about communication stacks, related security standards and challenges for emerging technologies such as LTE-V2X.

We highlight that while prior work has covered a wide range of the security and privacy aspects, most of the previous surveys focus on some of the issues and do not provide broad view of the field. We believe our work complements prior surveys and provides a holistic overview of existing V2X security issues and possible countermeasures.



\section{Conclusion}
\label{sec:conclusion}


In the near future V2X communication technology is expected to revolutionize the modern ground transportation system. With the emergence of this modern technology, V2X applications will potentially be targeted by the malicious entities (as evident by the recent real-world attacks on automotive systems~\cite{greenberg2015hackers,checkoway2011comprehensive,koscher2010experimental,petit2015remote}) and there is a requirement of layered defence mechanism to improve the resiliency of such systems. In this survey we provided an overview of current V2X security standards, potential security threats and different detection approaches. While in this paper our primary focus in  on V2X technology, the novel security mechanisms developed for V2X applications can be used to improve the security of broader safety-critical cyber-physical domains~\cite{humayed2017cyber,konstantinou2015cyber}. We believe this research will be tangential and valuable to the academic/industry researchers, developers, systems engineers and standardization agencies working in systems security fields in general.


\bibliographystyle{IEEEtran}

\bibliography{references}



\end{document}